\documentclass[lettersize,journal,onecolumn]{IEEEtran}
\usepackage{amsmath,amssymb,amsfonts}
\usepackage{algorithmic}
\usepackage{algorithm}
\usepackage{array}
\usepackage[caption=false,font=normalsize,labelfont=sf,textfont=sf]{subfig}
\usepackage{textcomp}
\usepackage{stfloats}
\usepackage{url}
\usepackage{verbatim}
\usepackage{graphicx}
\usepackage{cite}

\usepackage{siunitx} 
\usepackage{multirow} 
\usepackage[version=4]{mhchem} 
\usepackage{xcolor}

\begin{document}

\title{Optimizations for a Current-Controlled Memristor-based Neuromorphic Synapse Design} 

\author{Hritom Das,\IEEEmembership{ Member,~IEEE}, Rocco D. Febbo, Charles P. Rizzo, Nishith N. Chakraborty, \IEEEmembership{ Graduate Student Member,~IEEE},  James S. Plank~\IEEEmembership{Senior Member,~IEEE}, Garrett S. Rose~\IEEEmembership{Senior Member,~IEEE} 
\thanks{Manuscript received 9 April 2023; revised 22 June 2023; accepted 30 August 2023. This material includes research sponsored by the Air Force Research Laboratory under agreement number FA8750-21-1-1018. The U.S. Government may reproduce and distribute reprints for Governmental purposes, despite any copyright notation. The views and conclusions expressed herein are solely those of the authors and do not necessarily reflect the official policies or endorsements of the Air Force Research Laboratory or the U.S. Government. The authors are thankful to Ryan Weiss for his support in circuit design.  (Corresponding author: Hritom Das.)}  
\thanks{The authors are with the Department of Electrical Engineering and Computer Science, University of Tennessee, Knoxville, TN 37996, USA (e-mail: hdas, rfebbo, crizzo, nchakra1, jplank, garose@utk.edu)}}

\markboth{}%
{Shell \MakeLowercase{\textit{et al.}}: A Sample Article Using IEEEtran.cls for IEEE Journals}


\maketitle

\begin{abstract}
The synapse is a key element of neuromorphic computing in terms of efficiency and accuracy. In this paper, an optimized current-controlled memristive synapse circuit is proposed. Our proposed synapse demonstrates reliability in the face of process variation and the inherent stochastic behavior of memristors. Up to an 82\% energy optimization can be seen during the \textit{SET} operation over prior work. In addition, the \textit{READ} process shows up to 54\% energy savings. Our current-controlled approach also provides more reliable programming over traditional programming methods. This design is demonstrated with a 4-bit memory precision configuration.  
Using a spiking neural network (SNN), a neuromorphic application analysis was performed with this precision configuration. Our optimized design showed up to a 82\% improvement in control applications and a 2.7x improvement in classification applications compared with other design cases.  

\end{abstract}


\section{Introduction}
\IEEEPARstart {P}{erformance} scaling has always been a major focus in the silicon industry. However, along with increasing difficulty, the performance benefits of technology node downsizing have become significantly diminished. Thus, a search for new computer architectures beyond the classic von-Neumann architecture has been gaining popularity. One of these new architectures is in-memory computing which integrates computational circuits directly with memory to avoid memory transfer bottlenecks. In this architecture, new memory devices are being explored as well. These emerging memory devices offer power, performance, and area benefits when compared to conventional memory devices such as SRAM or DRAM \cite{r12, r13, r14}. One of the most popular candidates among emerging memory devices is the memristor\cite{r1}. Memristors are two terminal devices that can be programmed into a variable resistance. In addition to the low power and low area, they also operate in a non-volitile nature unlike common cache memories. Often, the non-Von-Neumann architecture used with memristors is a neuromorphic architecture with an in-memory or near-memory computing centric design \cite{r15, r16, r17, r18}.

The goal of neuromorphic computing is to mimic the highly efficient brain behavior observed in neuroscience. Fully deployable neuromorphic systems, commonly referred to as neuroprocessors, provide interfaces for a user to perform specific tasks while utilizing neuromorphic hardware. One of the ways neuroprocessor implementations can vary is by the type of internal signals such as digital, analog, or mixed. There are several different neuroprocessors available such as Loihi by Intel \cite{r2}, TrueNorth by IBM \cite{r3}, and SpiNNaker from The University of Manchester \cite{r4,r5}. All of which are digital implementations. BrainScaleS is a mixed-signal implementation \cite{r7}. RAVENS is a neuroprocessor standard which has an FPGA implementation as well as a mixed-signal memristor based implementation in development\cite{r6}. Neuroprocessor implementations can also vary by the type and implementation of neurons and synapses. The synapse, a key component which contains the weighted connections between neurons, can be implemented using an analog or digital design \cite{r6}. Neuroprocessor hardware design is inspired from biological structures where one neuron can have several thousand connections to other neurons via synapses. This creates several challenges for implementation in hardware. One of these challenges is managing the area required to store the weighted connection between neurons. Since memristors are very small, this challenge can be alleviated.

Using memristors, a single synapse can be implemented with a 1T1R structure where one transistor(T) provides access control to the memristor(R). Our approach proposes a 3T1R where an additional PMOS transistor isolates the memristor from $VDD$ during standby operation. Also, during the \textit{READ} operation, an NMOS transistor forms a second stage \textit{READ} current which significantly reduces the overall \textit{READ} power. This second stage current also allows for a constant current through the memristor which can improve reliability.

Here, hafnium oxide based memristors are considered for the synapse circuit design, which has suitable characteristics for analog memory \cite{beckmann2020towards}. There are four operations that can be performed on a memristor: \textit{FORM}, \textit{RESET}, \textit{SET}, and \textit{READ}. Several considerations must be made while performing these operations. The \textit{FORM} operation is a one-time initialization step. To do a forming operation with our proposed design a $PMOS$ will be connected to the top of the memristor. The top $PMOS$ and a $NMOS$ at the bottom of the memristor will be digitally “ON” during a \textit{FORM} operation. Here the \textit{FORM} $VDD$ is \SI{3.3}{\volt}. During a \textit{FORM} operation, there will be a filament creation from the top electrode to the bottom electrode. Once formed, the \textit{RESET} operation puts the memristor into a high resistance state (HRS) of a few hundred K$\Omega$ and must take place before a \textit{SET} operation. Next, the \textit{SET} operation puts the memristor into a low resistance state (LRS). For the \textit{SET} operation, a small range of LRS values from single K$\Omega$s to a few tens of K$\Omega$s can be targeted \cite{liehr2019fabrication, ryan}. While utilizing the HRS has benefits for implementing low power and a highly dynamic synapse \cite{Programming_Scheme}, the inherent stochastic behavior of memristors in a HRS can be mitigated by strictly using a LRS. Finally, for \textit{READ}, since the memristor is only a two-terminal device, care must be taken not to \textit{SET} or \textit{RESET} the device.

Unlike binary memory applications which would greatly benefit from separating the maximum possible resistance and minimum possible resistance \cite{binaryMemristor}, this synapse design uses the programmable LRS region mentioned before for implementing an analog memory. During \textit{RESET}, the HRS of these devices is reached by applying a voltage with the opposite polarity from the initial forming voltage. This switching method can achieve specific resistance values by applying voltage pulses. The main issues with this pulse-based approach are the sensitivities to process variation and tight requirements for the shape of the pulse\cite{r19}. Another dynamic switching method is performed by simply limiting the current during the \textit{SET} operation. This works due to Ohm's law. Since the resistance is decreasing and the current is constant, the lower-bound voltage threshold, which halts the \textit{SET} operation, will be reached based on the current applied. By varying the current applied during the \textit{SET} operation, various LRSs can be obtained.

 During programming, without a proper $VDD$\_$SET$ and transistor sizing, the available programming range can become saturated for 3T1R. This is due to a non-linearity in set current. Susceptibility to process variation must be minimized for \textit{SET} operation to enhance the reliability of programming the synapse and to obtain similar behavior across all synapses. In addition, the $VDD$ and transistor sizing for a \textit{READ} operation is important to achieve stability and low-power operation. Transistor and $VDD$ scaling is a popular approach to optimize traditional memory design \cite{r12, r13, r14, r20}. In this work, different sizing corners and $VDD$ are taken into consideration for \textit{SET} and \textit{READ} operations. According to our best knowledge, this is the first documented approach to optimize memristor-based current-controlled synapse design.           
 
The key contributions of this paper are as follows.  
\begin{enumerate}
    \item Current compliance synapse design with \ce{HfO2} based 1T1R memristor. The current compliance feature will limit the $1^{st}$ stage current and reduce the overall power and energy consumption of the circuit. 
    \item   Reliability of the \textit{SET} and \textit{READ} operation at LRS are illustrated. Monte Carlo simulations are utilized to observe the process variation of the proposed synapse and found an optimal design region with reliability. 
    \item	A low-power design with optimized sizing is presented. \textit{READ} current is optimized with \textit{READ} device scaling. \textit{READ} device scaling shows a significant amount of energy optimization with proper functionality.   
    \item	Distinguishable \textit{READ} current levels can be sensed for 4-bit  memory precision. A moderate amount of \textit{READ} current is illustrated for 4-bit memory precision.
    \item Proposed circuitry utilized for classification and control applications using a neuromorphic framework to observe the impact of our optimized design in deployment.
\end{enumerate}

The rest of the paper is organized as follows. Section II provides a brief background about our proposed synapse with hafnium oxide-based memristor device and the write-read operation. Section III shows the simulation and measured results. Section IV will illustrate usage of our proposed design in neuromorphic applications. This section exhibits the performance and power consumption of our synapse performing in classification and control applications. A comparison with prior works will be analyzed in Section V. Finally, the paper will be concluded with indications of future work in Section VI.
  
\begin{figure}[]
            \centering
            \includegraphics[width=3.4in]{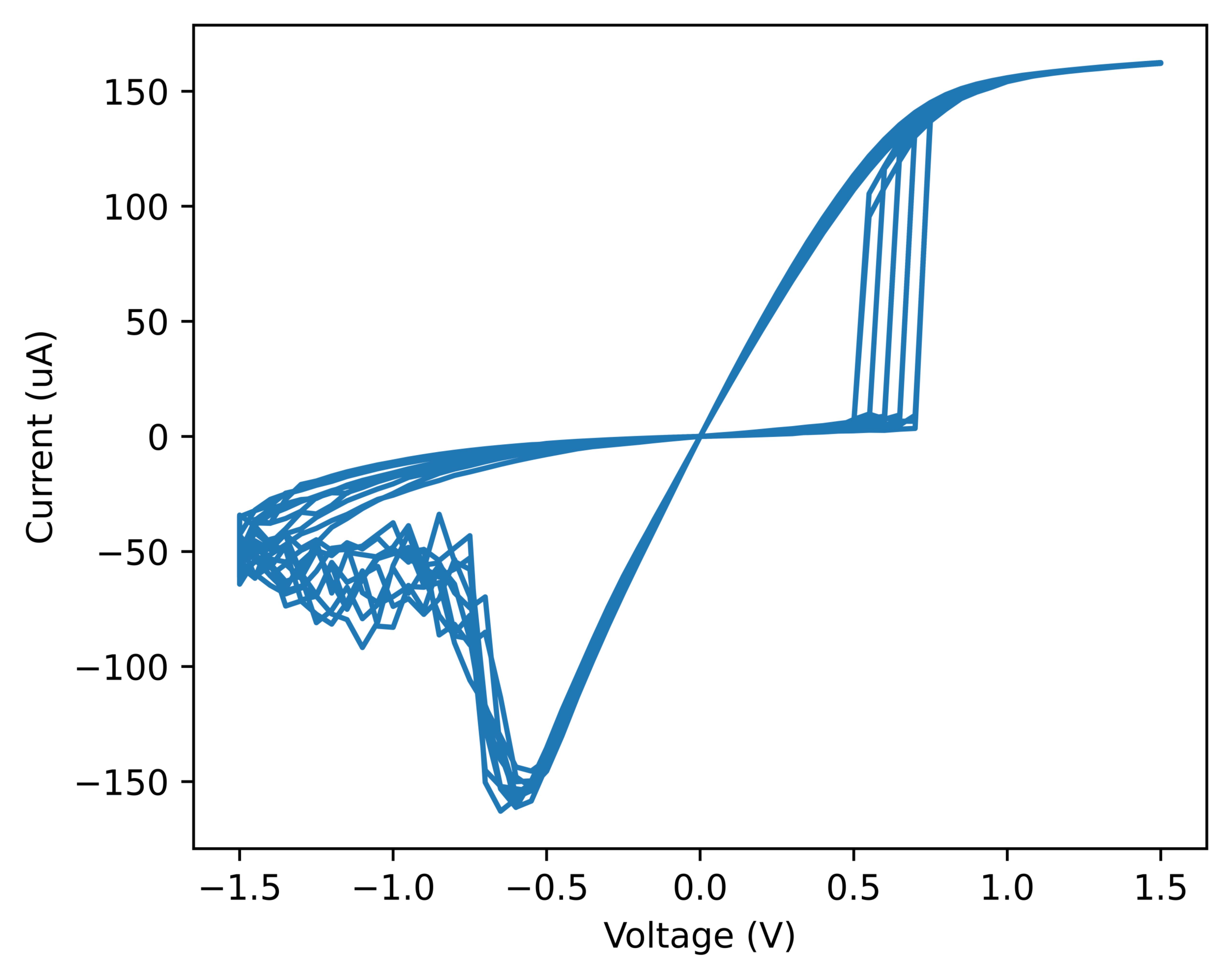}
            \caption{ Shows the I-V characteristics graph of a \ce{HfO2} based memristive device. }
            \label{fig:IV}
        \end{figure}
\section{Memristor based 3T1R synapse}
A \ce {TiN TE}\--\ce{HfO2}\--\ce{TiN BE} material stack is utilized to construct our memristor device. \ce{HfO2} is used in between the top electrode (TE) and bottom electrode (BE). In Fig. \ref{fig:IV} the device can be seen switching between two resistance levels (LRS and HRS). By limiting the current during the switch from a high resistance to a low resistance, a specific resistance level can be targeted. With this technique, an analog memory device can be designed. A positive current is observed for \textit{SET} operation and a negative current is illustrated for \textit{RESET} operation\cite{model_glsvlsi}. 

We collected measured results from fabricated memristor devices that use the \ce{HfO2} material. Based on the measured results such as I-V characteristics, we derived our Verilog-A model \cite{model_glsvlsi}. In our model, there are different measured and curve fitting parameters such as threshold voltages of LRS and HRS. The times required for different operations or switching are also included. We utilized a \SI{65}{\nano\meter} CMOS 10LPe process from IBM \cite{65}. Thick oxide transistors are utilized due to the high-forming voltage. Due to that a low-power design is difficult to achieve. 

\begin{figure}[]
            \centering
            \includegraphics[width=3.4in]{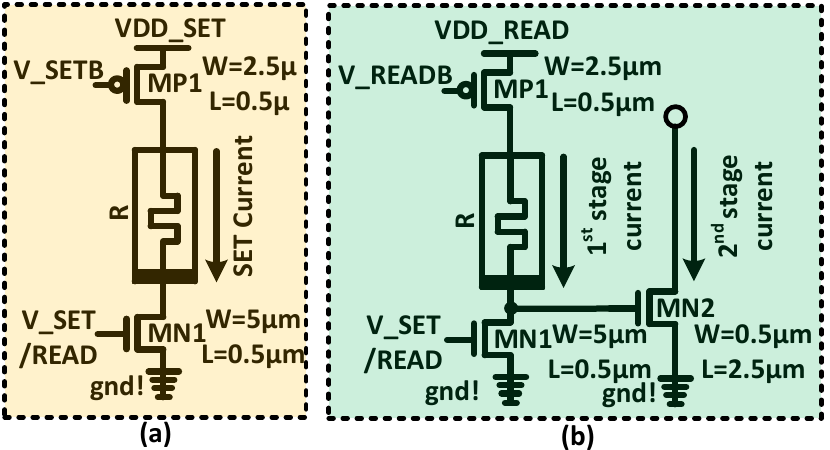}
            \caption{(a) shows the active transistors during a \textit{SET} operation. (b) shows the components used during the \textit{READ} operation. $M_{N1}$ and $M_{P1}$ devices are shared for \textit{SET} and \textit{READ} operation. Only the $2^{nd}$ stage current is considered during the \textit{READ} operation}
            \label{fig:synapse}
        \end{figure}

Fig. \ref{fig:synapse} illustrates the proposed 3T1R based memristive synapse. In (a), the circuitry required for the \textit{SET} operation is shown. In (b), the circuitry for the \textit{READ} operation is shown. This circuit is designed to utilize the memristor with considerations for the highly stochastic memristors while minimizing area and power. For the hafnium oxide memristor, transistors that are directly connected to the device need to be thick oxide in order to operate in a high-voltage region for forming. Thick oxide transistors are also useful to reduce flicker noise\cite{r10} which is useful in analog design. These large transistors are reused to control the memristor through various operations. The \textit{SET} operation utilizes $M_{N1}$ and $M_{P1}$. The \textit{READ} operation uses $M_{N1}$, $M_{N2}$, and $M_{P1}$. The transistors $M_{N1}$ and $M_{P1}$ are shared for the \textit{SET} and \textit{READ} operation. When programming a memristor in the linear region, it is important to be aware of the effects of process variation. This is done in order to avoid a case where two memristors are attempted to be programmed to the same value, but are not physically represented as the same value. A secondary objective of the \textit{READ} transistor is to minimize its width in order to perform with the lowest power since overall current is reduced. At the same time, it needs to be immune to process variation. In the next subsections, \textit{SET} and \textit{READ} operations will be discussed in detail with a focus on sizing, power, reliability, and process variation effects.

\subsection{Programming / SET}
Our proposed \textit{SET} circuitry is designed with sizing optimized to perform reliable programming at low power. Fig. \ref{fig:synapse} (a) shows the circuitry utilized during a \textit{SET} operation with one memristor and two transistors $M_{N1}$ and $M_{P1}$. The \textit{SET} operation utilizes \SI{3.3}{\volt} as a $VDD$ to program the memristor to an LRS anywhere from 5 k$\Omega$ to 20 k$\Omega$, which is more linear and reliable than an HRS. Before the \textit{SET} operation, a \textit{RESET} operation is needed. Once the device is reset into an HRS at 100  k$\Omega$, it is programmed into an LRS with the \textit{SET} operation. During the \textit{SET} operation, $M_{N1}$ is used to limit the current through the device by using different voltages at the gate. The current produced when $M_{N1}$ is in saturation results in different resistances for the memristor.

\begin{figure}[t]
            \centering
            \includegraphics[width=3.4in]{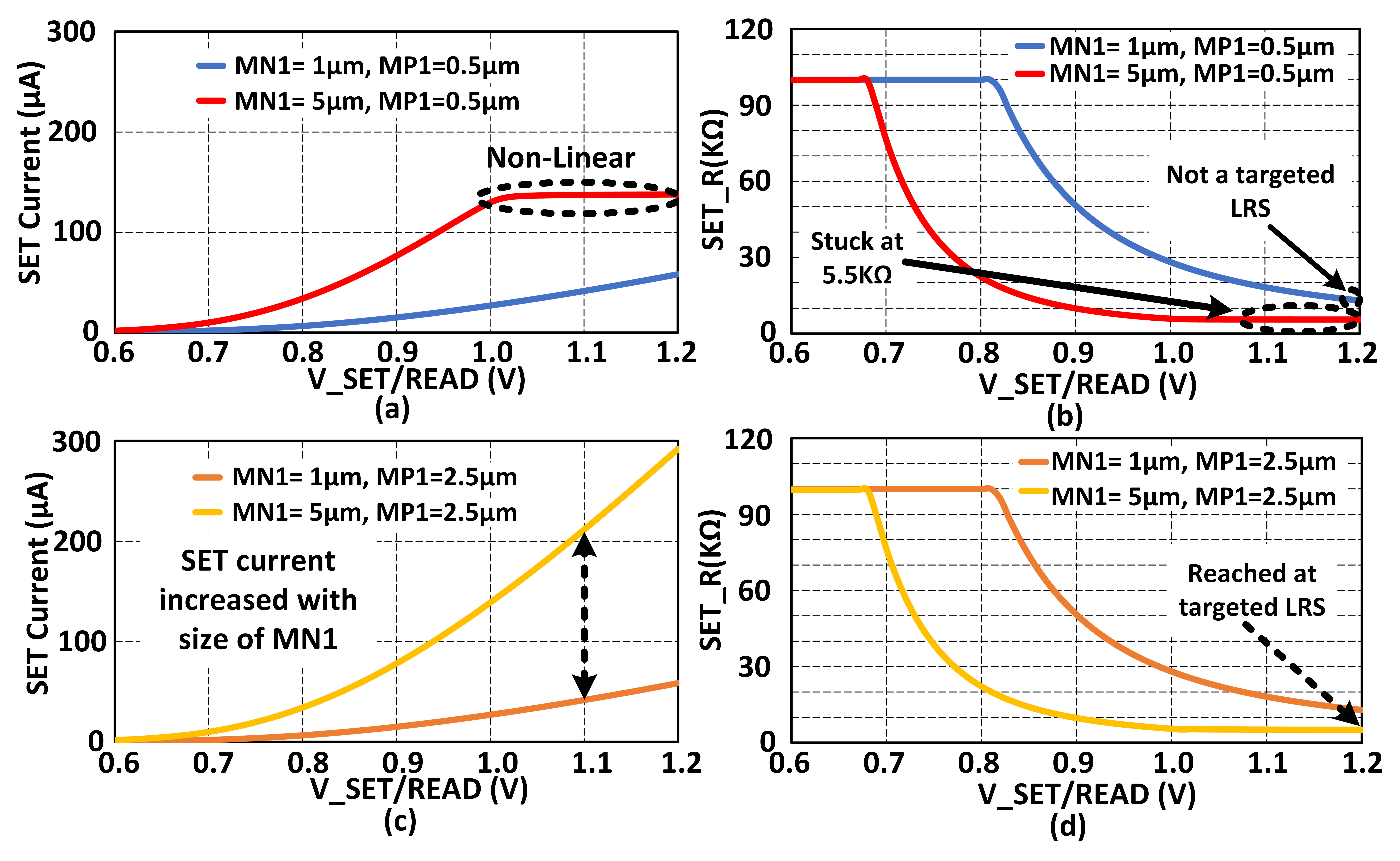}
            \caption{The simulation results of \textit{SET} currents and achievable resistance levels with different sizing corners. (a) and (b) shows the variability of \textit{SET} current and resistance while $M_{P1}$ is \SI{0.5}{\micro\meter}. (c) and (d) exhibit the current and resistance level when $M_{P1}$ and $M_{N1}$ are large enough to provide a linear operating region for \textit{SET} operation. If $M_{N1}$ is too small, the targeted LRS cannot be achieved.}
            \label{fig:SET_current_R}
        \end{figure}

Fig. \ref{fig:SET_current_R} shows the current and achievable resistance of the memristor during \textit{SET} operation. Here, four corner positions are considered to obtain optimal sizing of $M_{N1}$  and $M_{P1}$. Fig. \ref{fig:SET_current_R}(a), shows that increasing $M_{N1}$s width will increase the \textit{SET} current. However, a saturation region appears when the width of $M_{N1}$ is increased. This effectively limits the available programming region as seen in Fig. \ref{fig:SET_current_R}(b). In Fig. \ref{fig:SET_current_R}(c) when the width of $M_{P1}$ and $M_{N1}$ are increased, the overall \textit{SET} current is increased without the non-linear effects. The effects of this increase in current can be seen in part (d) of Fig. \ref{fig:SET_current_R}. With an increase of both widths, a higher dynamic range of resistances can be targeted during the \textit{SET} operation. Due to that, larger $M_{N1}$ and $M_{P1}$ widths is better for reliability at the cost of higher current dissipation. This synapse is designed to have a gate voltage range of \SI{0.8}{\volt} to \SI{1.2}{\volt} at $M_{N1}$, so it can be driven using smaller, lower voltage transistors. These gate voltages result in saturation currents near \SI{34.3}{\micro\ampere} to \SI{291.8}{\micro\ampere} as seen in Fig. \ref{fig:SET_current_R}(c). The p-type \textit{SET} transistor, $M_{P1}$, provides the high voltage $VDD$\_$SET$ necessary to create the filament during forming. Thus, $VDD$\_$SET$ is set to \SI{3.3}{\volt}. In this method, the resistance of the device is reliably programmable via current control, providing a reliable resistance range from 5 to \SI{20}{\kilo\ohm}.

\begin{figure}[]
            \centering
            \includegraphics[width=3.4in]{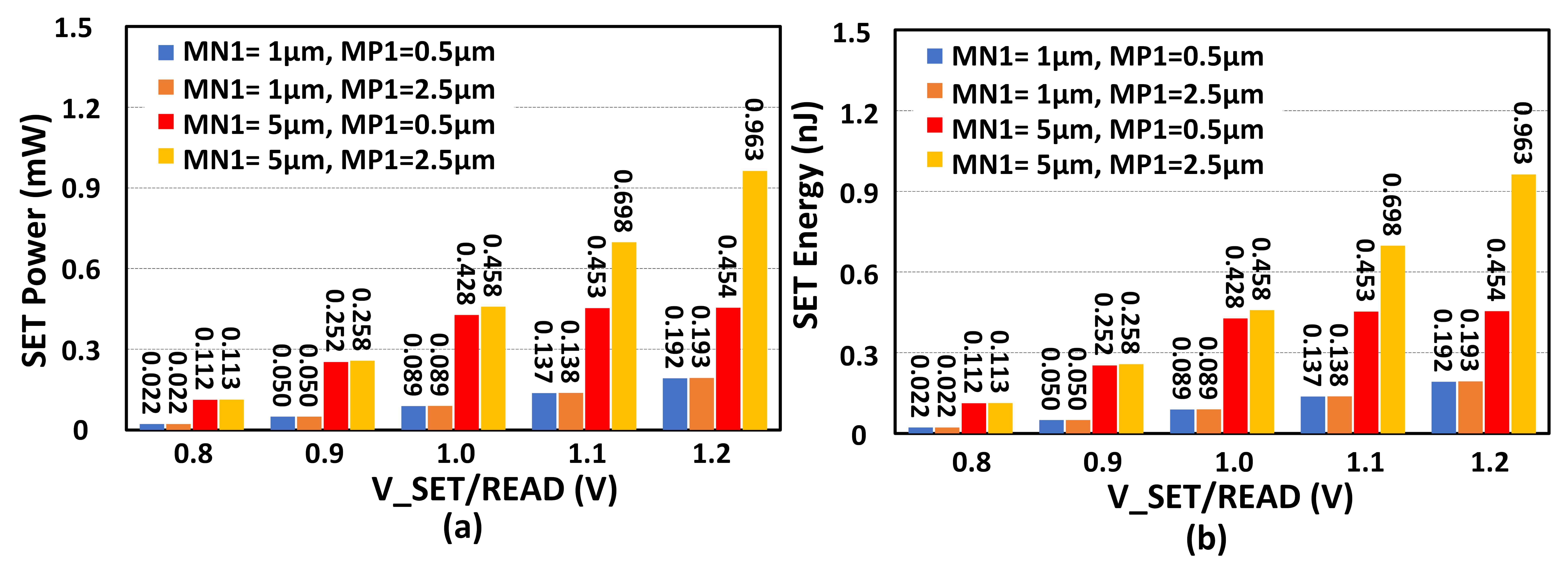}
            \caption{Simulation results for \textit{SET} power and energy across the different configurations. (a) shows the \textit{SET} power consumption at different \textit{SET} voltages. (b) illustrates the \textit{SET} energy for \SI{1}{\micro\sec} window.}
            \label{fig:SET_Power and Energy}
        \end{figure}

Fig. \ref{fig:SET_Power and Energy}, shows the power and energy consumption with our proposed design at different programming voltages and transistor sizings. All simulations are performed with a Verilog-A memristor model\cite{model_glsvlsi, ryan} and 65nm CMOS technology. Fig. \ref{fig:SET_Power and Energy} (a), shows the power consumption of four different sizing combinations. At \SI{0.8}{\volt}, the \textit{SET} current for the largest sizing configuration is about \SI{0.113}{\milli\watt}. The resulting resistance with $V\_SET\slash READ$ at \SI{0.8}{\volt} is \SI{20}{\kilo\ohm}. In addition, at \SI{1.2}{\volt}, the achievable \textit{SET} resistance is about \SI{5}{\kilo\ohm} with \SI{0.963}{\milli\watt}. A \SI{1}{\micro\sec} window is considered for energy consumption. The energy of our proposed design is \SI{0.113}{\nano\J} and \SI{0.963}{\nano\J} at \SI{0.8}{\volt} and \SI{1.2}{\volt} respectively. To achieve a reliable and targeted LRS, higher power and energy are considered. Higher $VDD$ and thick oxide transistors are also responsible for higher power and energy consumption.                     

        \begin{figure}[t]
            \centering
            \includegraphics[width=3.4in]{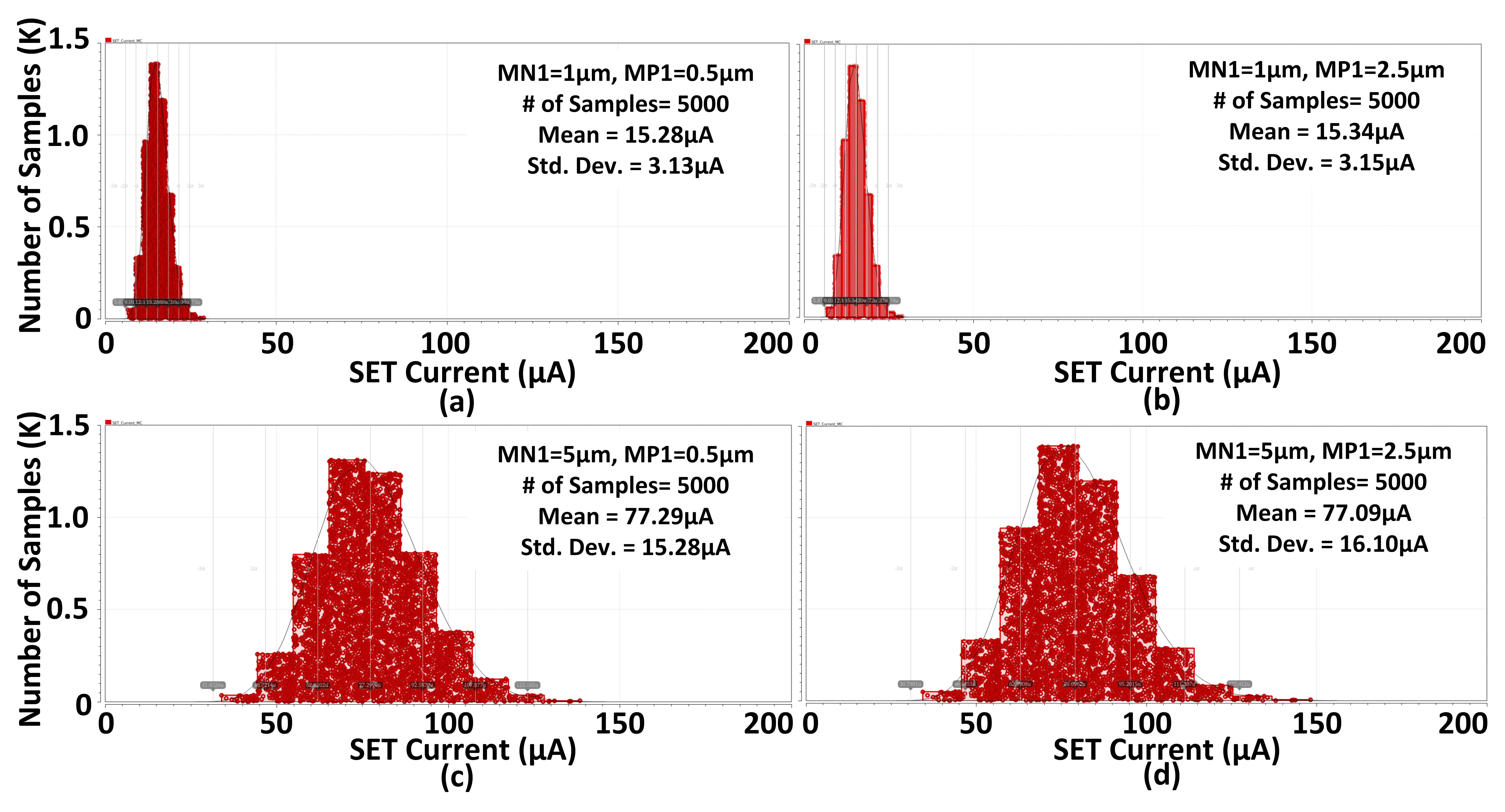}
            \caption{Monte Carlo simulation runs on Cadence at 65nm CMOS process for the \textit{SET} operation. All four sizing corners are taken under consideration with 5000 Sample points. (a) and (b) shows the variability if $M_{N1}$s width fixed at \SI{1}{\micro\meter} and $M_{P1}$s width gets scaled from \SI{0.5}{\micro\meter} to \SI{2.5}{\micro\meter}. Next, (c) and (d) show the variability of the memristors \textit{SET} current when $M_{N1}$s width is fixed at \SI{5}{\micro\meter} and $M_{P1}$s width is scaled from \SI{0.5}{\micro\meter} to \SI{2.5}{\micro\meter}. \textit{SET} voltage is \SI{0.9}{\volt} for all MC simulations.}
            \label{fig:SET_MC}
        \end{figure}

According to Fig. \ref{fig:SET_MC}, Monte Carlo simulations in all four sizing corners vary in susceptibility to process variation. Simulations were run using Cadence Virtuoso on a 65nm CMOS process with 5000 sample points to observe the reliability of \textit{SET} current. Fig. \ref{fig:SET_MC} (a) and (b) show the mean and std. dev. is lower with smaller width of $M_{N1}$. On the other hand, Fig. \ref{fig:SET_MC} (c) and (d), show the mean and std. dev. are larger with the larger size of $M_{N1}$. Considering the ratio between the mean and std. dev. of the designs, larger transistors will help ensure a more reliable \textit{SET} operation. 
Due to that, we would like to move forward with the size of $M_{N1}$ and $M_{P1}$ are \SI{5}{\micro\meter} and \SI{2.5}{\micro\meter}. In this way, we can make sure the \textit{SET} operation is reliable and compensation for process variation is made.

\subsection {READ}

According to Fig. \ref{fig:synapse} (b), the read-out process uses $M_{N2}$ to create an output current ($2^{nd}$ stage current) based on the voltage at the bottom node of the memristor, which is induced by $1^{st}$ stage current. The memristor in its low resistance state is supplied a current through $M_{N1}$. This current is as low as possible while still being able to generate a reliable output.  
With \SI{0.6}{\volt} at the gate of $M_{N1}$, approximately \SI{1.8}{\micro\ampere} of current is generated. The p-type transistor $M_{P1}$ is selected when V\_READB is \SI{0}{\volt}. This $1^{st}$ stage current passes through the resistances of the memristor and $M_{P1}$ and sets the voltage at the gate of $M_{N2}$. The expected voltage is near \SI{0.6}{\volt} depending on the resistance in the memristor. If the voltage is higher more current will pass through the $2^{nd}$ stage current. 

\begin{figure}[]
            \centering
            \includegraphics[width=3.4in]{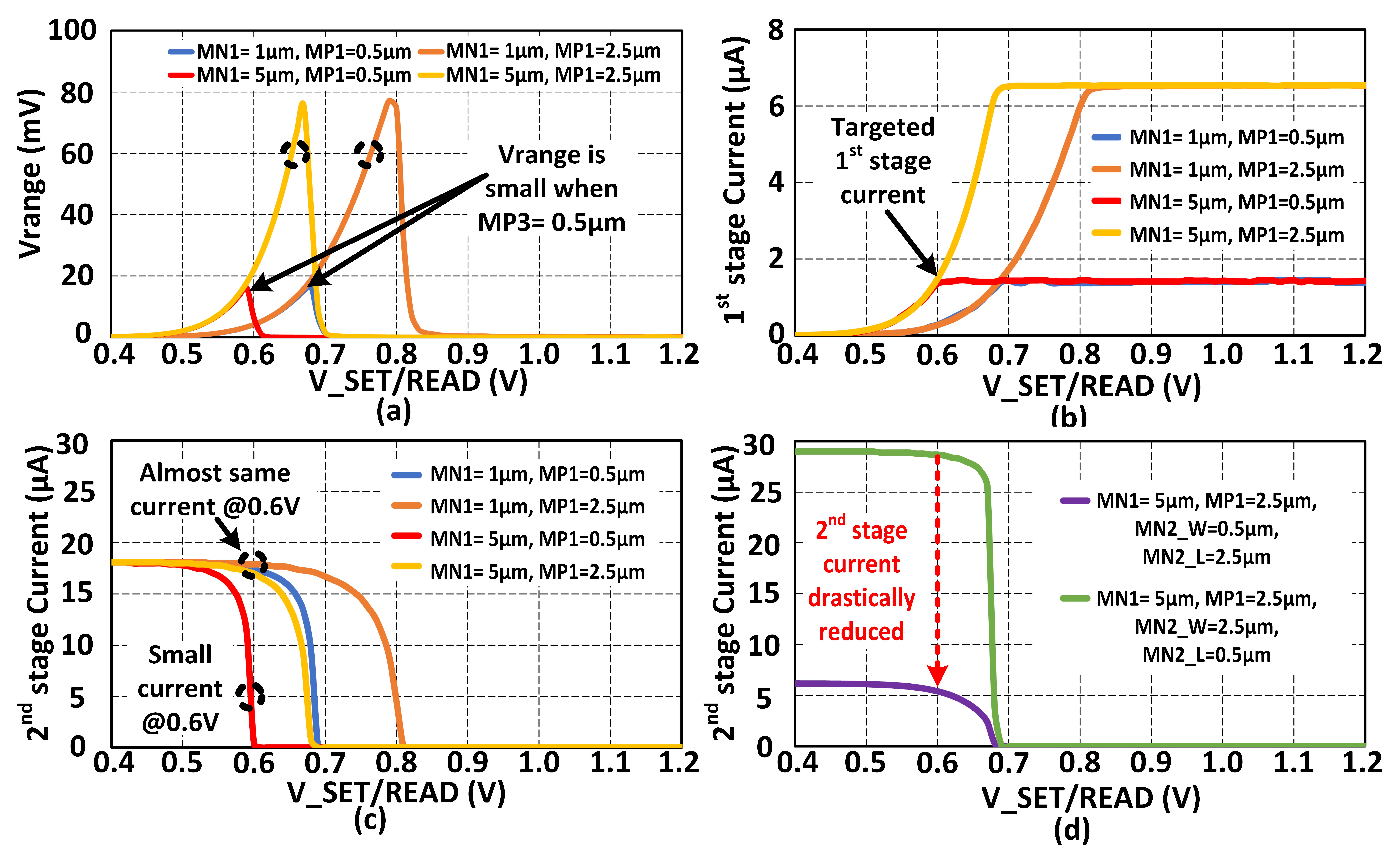}
            \caption{\textit{READ} simulation results are illustrated based on sizing of \textit{READ} devices $M_{P1}$, $M_{N1}$, and $M_{N2}$. (a) shows the relationship between voltage range (Vrange) and V\_SET\slash READ. The window of voltage range can be increased with the increment of V\_SET\slash READ. Also, different sizing shows different voltage ranges of \textit{READ} operation. (b) shows the $1^{st}$ stage current vs. V\_SET\slash READ. The relation among V\_SET\slash READ, 1st stage current and sizing of $M_{N1}$ and $M_{P1}$ are shown in (b). (c) illustrates the $2^{nd}$ stage current of the synapse that is enabled by the drain voltage of $M_{N1}$. Here, $M_{N2}$ kept \SI{0.5}{\micro\meter} for both width and length. The current of $2^{nd}$ stage is much larger than the $1^{st}$ stage. To reduce $2^{nd}$ stage current the length of the $M_{N2}$ increased. (d) shows the optimal design parameter to get a small current for $2^{nd}$ stage with a larger resistance of $M_{N2}$.}
            \label{fig:Vnset_Vrange}
        \end{figure}

Fig. \ref{fig:Vnset_Vrange} (a), shows the simulation results of V\_SET\slash READ vs. $V_{range}$. V\_SET\slash READ is the gate voltage of $M_{N1}$ and the $V_{range}$ is the difference between two drain voltages of $M_{N1}$ when the resistance is \SI{5}{\kilo\ohm} and \SI{20}{\kilo\ohm}. $V_{range}$ should be large in order for $M_{N2}$ to provide a large current swing in the $2_{nd}$ stage. At \SI{0.6}{\volt}, if $M_{P1}$ is \SI{0.5}{\micro\meter}, $V_{range}$ is too low. However, if $M_{P1}$ is \SI{2.5}{\micro\meter}, the voltage range moves from \SI{6}{\milli\volt} to \SI{22}{\milli\volt} depending on the width of $M_{N1}$. Moreover, if the V\_SET\slash READ increased from \SI{0.6}{\volt} to \SI{0.65}{\volt}, then the voltage range will vary from \SI{11}{\milli\volt} to \SI{57}{\milli\volt}. This provides a better voltage window for \textit{READ} operation. However, it will increase the \textit{READ} current of $1^{st}$ stage. When $M_{N1}$ and $M_{P1}$ are \SI{1}{\micro\meter} and \SI{2.5}{\micro\meter} respectively $V_{range}$ is \SI{57}{\milli\volt}. However, this caused the gate voltage to be increased to \SI{760}{\milli\volt}. If the gate voltage of $M_{N1}$ is increased too much, the stability of $2^{nd}$ stage current will be decreased. This also causes the $1^{st}$ stage current to increase, which is visible in Fig. \ref{fig:Vnset_Vrange} (b).

According to Fig. \ref{fig:Vnset_Vrange} (b), while reading at \SI{0.6}{\volt} the $1^{st}$ stage current is not influenced by the width of $M_{P1}$ when $M_{N1}$ is \SI{1}{\micro\meter} wide. Either of these could be a good design choice for low power and compact design. In that case, there would be a different $N_{MOS}$ for \textit{SET} and \textit{READ} operations since \textit{SET} operations require a larger $N_{MOS}$. To avoid this, $M_{N1}$ is increased to \SI{5}{\micro\meter}, and $M{P1}$ varies from \SI{0.5}{\micro\meter} to \SI{2.5}{\micro\meter}. In this configuration, $1^{st}$ stage current is increased from \SI{1.35}{\micro\ampere} to \SI{1.48}{\micro\ampere} at \SI{0.6}{\volt}. Here, $M_{N1}$ has a significant effect on the $1^{st}$ stage current. The sizing decision of $M_{N1}$ gives us the opportunity to eliminate one transistor from the synapse. Otherwise, we have to use two $N_{MOS}$ transistors, one for SET and another one for \textit{READ} operation, which will introduce an extra silicon area in the design.  

Now, consider the $2^{nd}$ stage current with the minimum size of $M_{N2}$ illustrated in Fig. \ref{fig:Vnset_Vrange} (c). Here, both the width and length of $M_{N2}$ are \SI{0.5}{\micro\meter}. In addition, the width of $M_{N1}$ and $M_{P1}$ are varied from \SI{1}{\micro\meter} to \SI{5}{\micro\meter} and \SI{0.5}{\micro\meter} to \SI{2.5}{\micro\meter} respectively. The current readings in  Fig. \ref{fig:Vnset_Vrange} (c), from about \SI{17}{\micro\ampere} to \SI{18}{\micro\ampere}, exceed the limit our neuron can operate on. Only the red curve shows a lower $2{nd}$ stage current, which is \SI{0.38}{\micro\ampere}. If this configuration shows stable \textit{READ} current with Monte Carlo simulation then it would be the option for low-power design. Hence, $M_{N1}$ and $M_{P1}$ are fixed at \SI{5}{\micro\meter} and \SI{2.5}{\micro\meter} respectively in \ref{fig:Vnset_Vrange} (d). In this figure, the effect of the length and width of $M_{N2}$ is analyzed to find an optimal solution for low power and stable operation. When the width of $M_{N2}$ is \SI{2.5}{\micro\meter} and length is \SI{0.5}{\micro\meter}, the $2{nd}$ stage output current increases to \SI{28.7}{\micro\ampere} at \SI{0.6}{\volt}. Due to the reduction in resistance, the current increases. Later, the width of $M_{N2}$ remains at \SI{0.5}{\micro\meter} with the length increased to \SI{2.5}{\micro\meter}. When the length of $M_{N2}$ is \SI{2.5}{\micro\meter} and the width is \SI{0.5}{\micro\meter}, the $2{nd}$ stage output current is reduced to \SI{5.4}{\micro\ampere} from \SI{28.7}{\micro\ampere}. As we know, the length of the transistor is proportional to resistance. Due to that, the $2{nd}$ stage output current is reduced with an area and latency overhead. $2{nd}$ stage current can be reduced further by increasing the length of $M_{N2}$.

\begin{figure}[]
            \centering
            \includegraphics[width=3.4in]{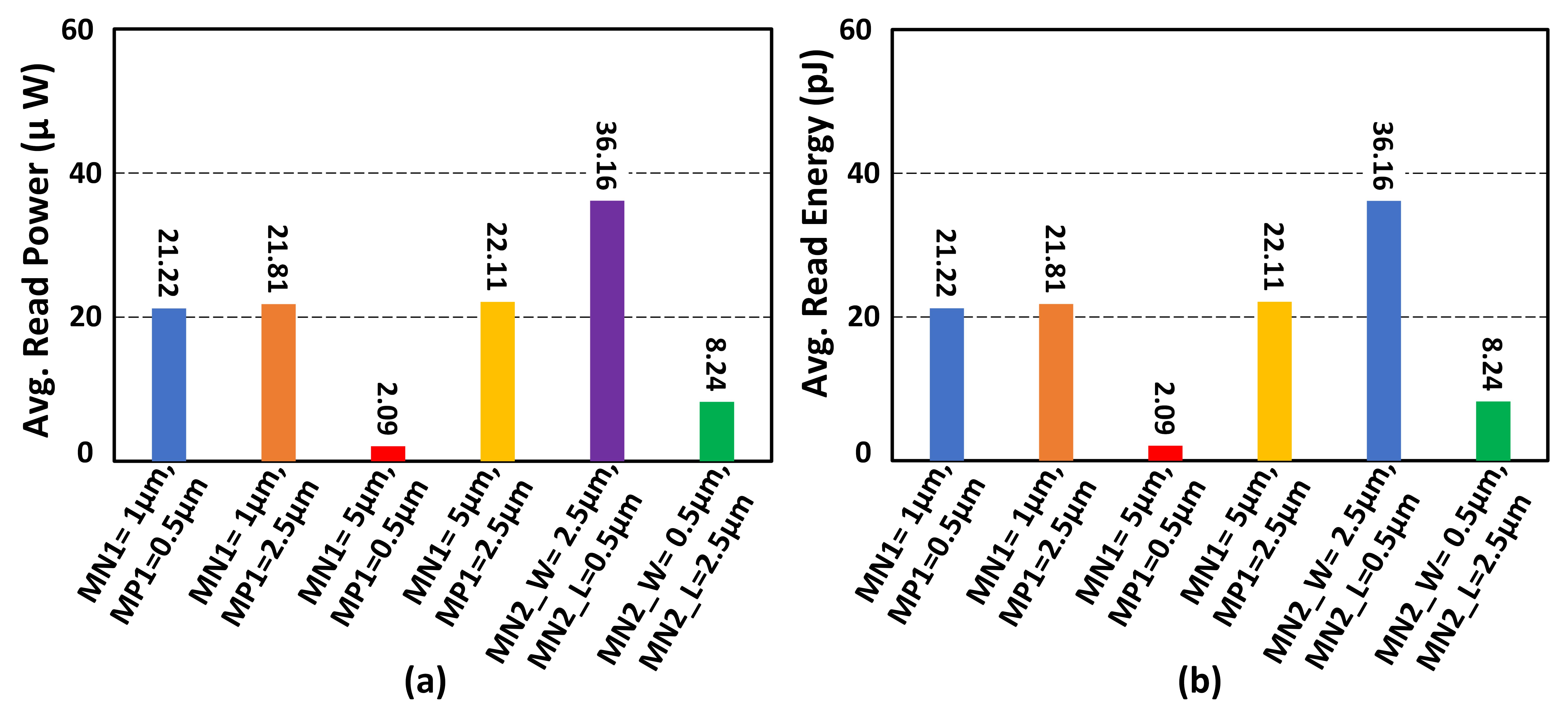}
            \caption{Simulation results for read power and energy. \textit{READ} operation conducted at \SI{0.6}{\volt}. (a) shows the \textit{READ} power chart for different sizing. Red and the green colored chart shows promising power consumption. (b) exhibits the \textit{READ} energy consumption for different device sizing. Every power and energy chart is averaged across \SI{5}{\kilo\ohm} to \SI{20}{\kilo\ohm} data.}
            \label{fig:Avg_Raed_PE}
        \end{figure}

Fig. \ref{fig:Avg_Raed_PE} shows the average power and energy consumption during a read operation. There are six different sizing combinations for simulations. Fig. \ref{fig:Avg_Raed_PE} (a), shows the power simulations for the read operation at each sizing and averaged across the resistance levels (integer) from \SI{5}{\kilo\ohm} to \SI{20}{\kilo\ohm}. When $M_{N1}$ is \SI{1}{\micro\meter}, the average power is about \SI{21}{\micro\watt}. If $M_{N1}$ is \SI{5}{\micro\meter} and $M_{P1}$ varies from \SI{0.5}{\micro\meter} to \SI{2.5}{\micro\meter}, the average power is varying from \SI{2.09}{\micro\watt} to \SI{22.11}{\micro\watt}. Finally $M_{N1}$ and $M_{P1}$ are kept at \SI{5}{\micro\meter} and \SI{2.5}{\micro\meter} and the length and width of $M_{N2}$ are varied. The average read power is \SI{8.24}{\micro\watt}, when the length of the $M_{N2}$ is \SI{2.5}{\micro\meter}. Fig. \ref{fig:Avg_Raed_PE} (b), shows the average energy of the read operation at different sizing. All the energies are at the \SI{}{\pico\joule} scale with a \SI{1}{\micro\s} clock which can be expected for an optimized synapse. The red and green chart shows optimized energy consumption for the \textit{READ} operation.

\begin{figure}[]
            \centering
            \includegraphics[width=3.4in]{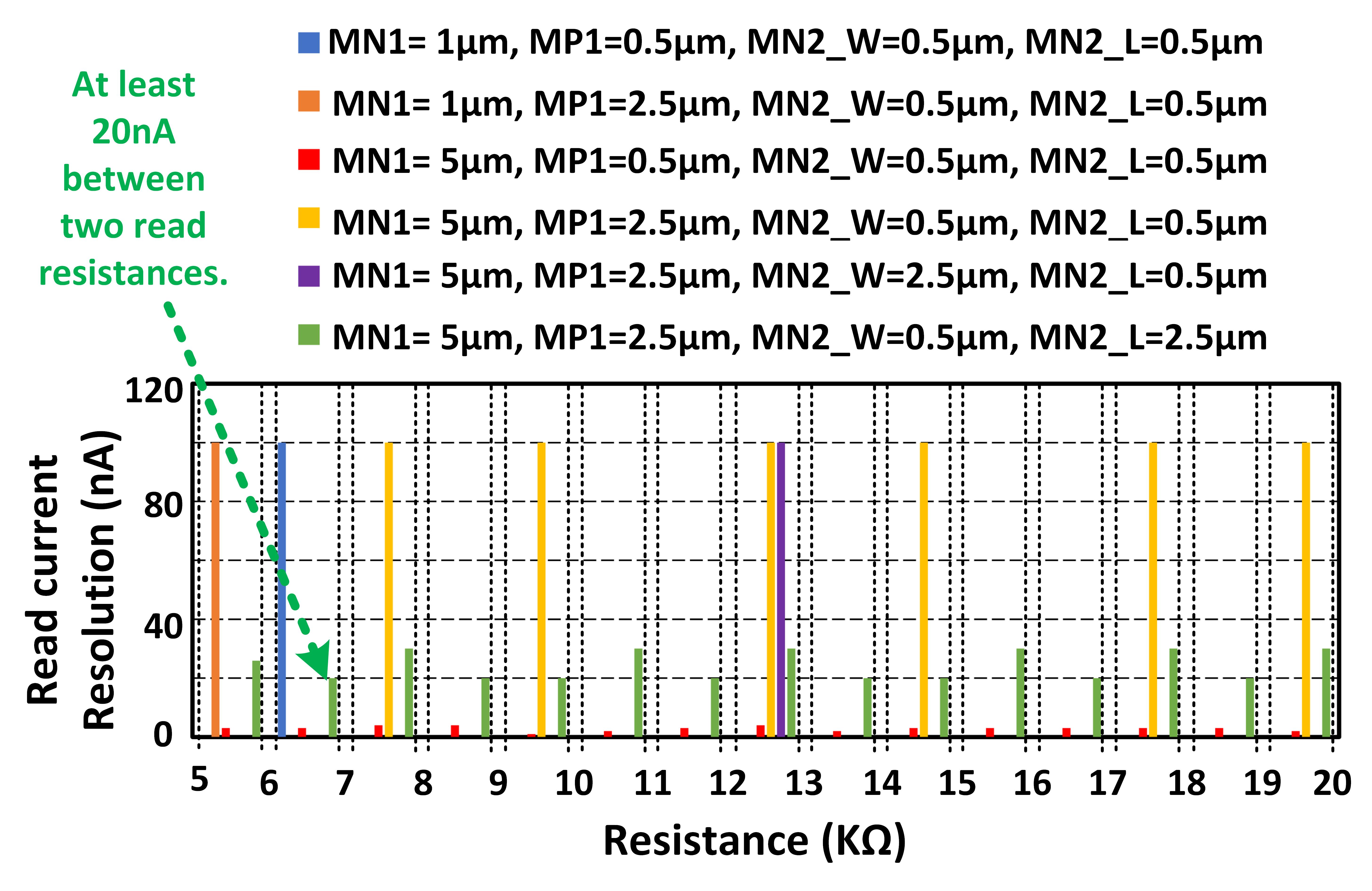}
            \caption{\textit{READ} current resolution is presented with simulation results. There are 16 different resistance levels to represent 4-bit data. All resistance levels draw a particular current to convert it from analog to digital data. Depending on transistor sizing, the \textit{READ} currents show different patterns at different resistance levels.}
            \label{fig:Read_I_Resolution}
        \end{figure}

Now, the resolution of $2^{nd}$ stage current is taken under consideration for multi-bit data management. Our ability to present 4-bit data with our targeted LRS range is dependent on the resolution of the $2^{nd}$ stage current. If the \textit{READ} current is not differentiable between resistance levels, then the analog to digital converter (ADC) can not differentiate all 16 combinations of data. Let's assume \SI{5}{\kilo\ohm} is "0000" and  \SI{20}{\kilo\ohm} is "1111". There should be a significant amplitude difference in the current between the two resistance levels. This significant amplitude difference is also important to any other circuit such as a neuron. According to Fig. \ref{fig:Read_I_Resolution}, when $M_{N1}$ is \SI{1}{\micro\meter}, the current difference between \SI{5}{\kilo\ohm} to \SI{6}{\kilo\ohm} and \SI{6}{\kilo\ohm} to \SI{7}{\kilo\ohm} is about \SI{100}{\nano\ampere}. However, his configuration has no stable pattern for the remaining resistance levels. When $M_{N1}$ and $M_{P1}$ are \SI{5}{\micro\meter} and \SI{0.5}{\micro\meter}, the current difference is very small, which would make it difficult for an ADC to detect. In addition, when $M_{N1}$ and $M_{P1}$ are \SI{5}{\micro\meter} and \SI{2.5}{\micro\meter}, and both length and width of $M_{N2}$ is \SI{0.5}{\micro\meter} the current difference is about \SI{100}{\nano\ampere}. Again it is not consistent for every resistance level. Finally, the green-colored data shows promising results for \textit{READ} current resolutions. Between two \textit{READ} currents, it shows at least \SI{20}{\nano\ampere} difference for all resistance levels. Finally, process variation resilience is analyzed using Monte Carlo simulations for the \textit{READ} currents in both stages.                   
 
 \begin{figure}[t]
            \centering
            \includegraphics[width=3.4in]{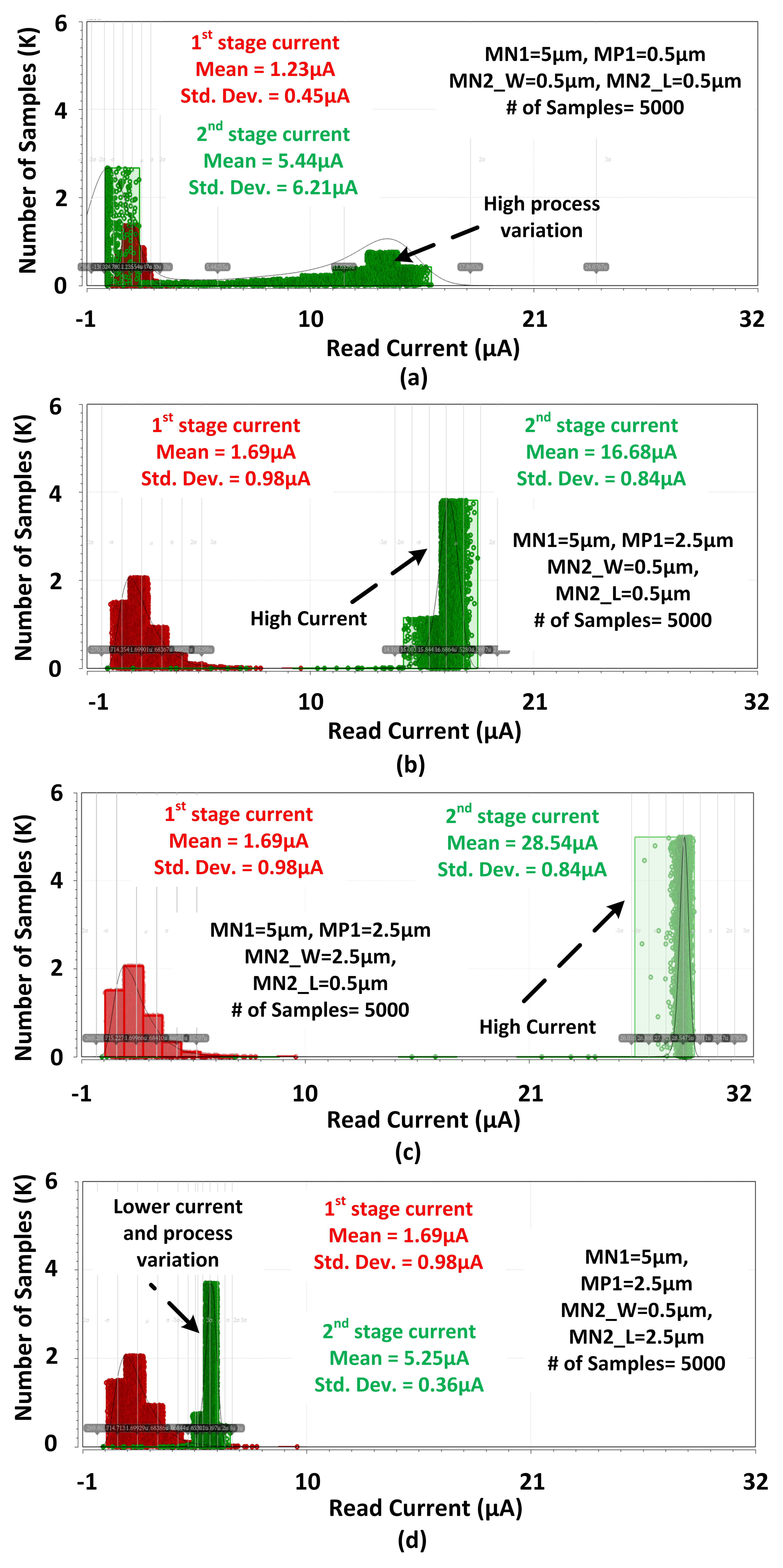}
            \caption{Monte Carlo simulation for \textit{READ} current with different sizes of $M_{P1}$, $M_{N1}$, and $M_{N2}$ transistors. (a) and (b) show the simulation results for the $1^{st}$ and $2^{nd}$ stage currents when the size of $M_{N1}$ and $M_{N2}$ are fixed at \SI{5}{\micro\meter} and \SI{0.5}{\micro\meter} respectively. In addition, the size of $M_{P1}$ varies from \SI{0.5}{\micro\meter} to \SI{2.5}{\micro\meter}. (c) and (d) illustrates the simulation results when both length and width of $M_{N2}$ vary from \SI{0.5}{\micro\meter} to \SI{2.5}{\micro\meter} and $M{N1}$ and $M{P1}$ are fixed at \SI{5}{\micro\meter} and \SI{2.5}{\micro\meter} respectively.}
            \label{fig:Raed_MC}
        \end{figure}
        
Reliability or process variation resilience has been measured based on the sizing of $M_{N1}$, $M{P1}$, and $M_{N2}$ with Monte Carlo simulations. Due to current compliance operation, $1^{st}$ stage current is always low. However, $2^{nd}$ stage current needs to be reduced for low-power design. Relying on low current stages can be difficult due to process variation. Process variation effects are observed in Fig. \ref{fig:Raed_MC} for all possible corners. According to Fig. \ref{fig:Raed_MC} (a) the mean and std. dev. of $2^{nd}$ stage current are \SI{5.44}{\micro\ampere} and \SI{6.21}{\micro\ampere}, which shows high sensitivity to process variation. Fig. \ref{fig:Raed_MC} (b) shows a higher \textit{READ} current with low std. dev. For low-power design, this design option is not viable due to the high current. Fig. \ref{fig:Raed_MC} (c) shows a higher mean current than (b) with the same std. dev. Finally, Fig. \ref{fig:Raed_MC} (d) exhibits a minimal mean and std. dev. among all design options. According to Fig. \ref{fig:Raed_MC} (d), the mean and std. dev. is \SI{5.25}{\micro\ampere} and \SI{0.36}{\micro\ampere} respectively. As a result, this design option can provide low power and a more reliable feature for \textit{READ} operation. Considering all analysis, it can be said that we will get an optimal design solution for SET and \textit{READ} operation when the size of the transistors will be as follows in Table \ref{tab:sizing}.  

The $PMOS$ ($MP_1$) can be shared in a column during standby operation. This scenario is considered with our neural core structure, where we design 16 synapses in a column and a $PMOS$ is utilized to control a column of synapses.

\begin{table}[]
\centering
    \caption{Transistor scaling for SET and READ operation}

\begin{tabular}{|c|c|c|}
\hline
Transistor Name & Width (µm) & Length (µm) \\ \hline
$M_{P1}$             & 2.5        & 0.5         \\ \hline
$M_{N1}$             & 5          & 0.5         \\ \hline
$M_{N2}$             & 0.5        & 2.5         \\ \hline
\end{tabular}%

 \label{tab:sizing}
\end{table}

\section{Results}
This section discusses simulation and measured results for process variations in the LRS and HRS using the sizing option shown in Table \ref{tab:sizing}. Cadence Virtuoso is used for simulation with a \SI{65}{\nano\meter} CMOS process and a Verilog-A memristor model. $VDD$ for \textit{SET} and \textit{READ} operations are \SI{3.3}{\volt} and \SI{1.2}{\volt} respectively. Measured results are observed from a chip fabricated using \SI{65}{\nano\meter} CMOS technology and hafnium-oxide as the memristor layer. Here, the forming operation was measured across 26 devices. These results show a mean forming voltage is \SI{1.66}{\volt} and a std. dev. is \SI{0.4}{\volt} which is reasonable for a reliable design. By limiting the current during a form, a more reliable resistance state can be achieved after forming. However, since the device will be subsequently RESET then SET, this is not an issue.    

\begin{figure}[]
            \centering
            \includegraphics[width=3.4in]{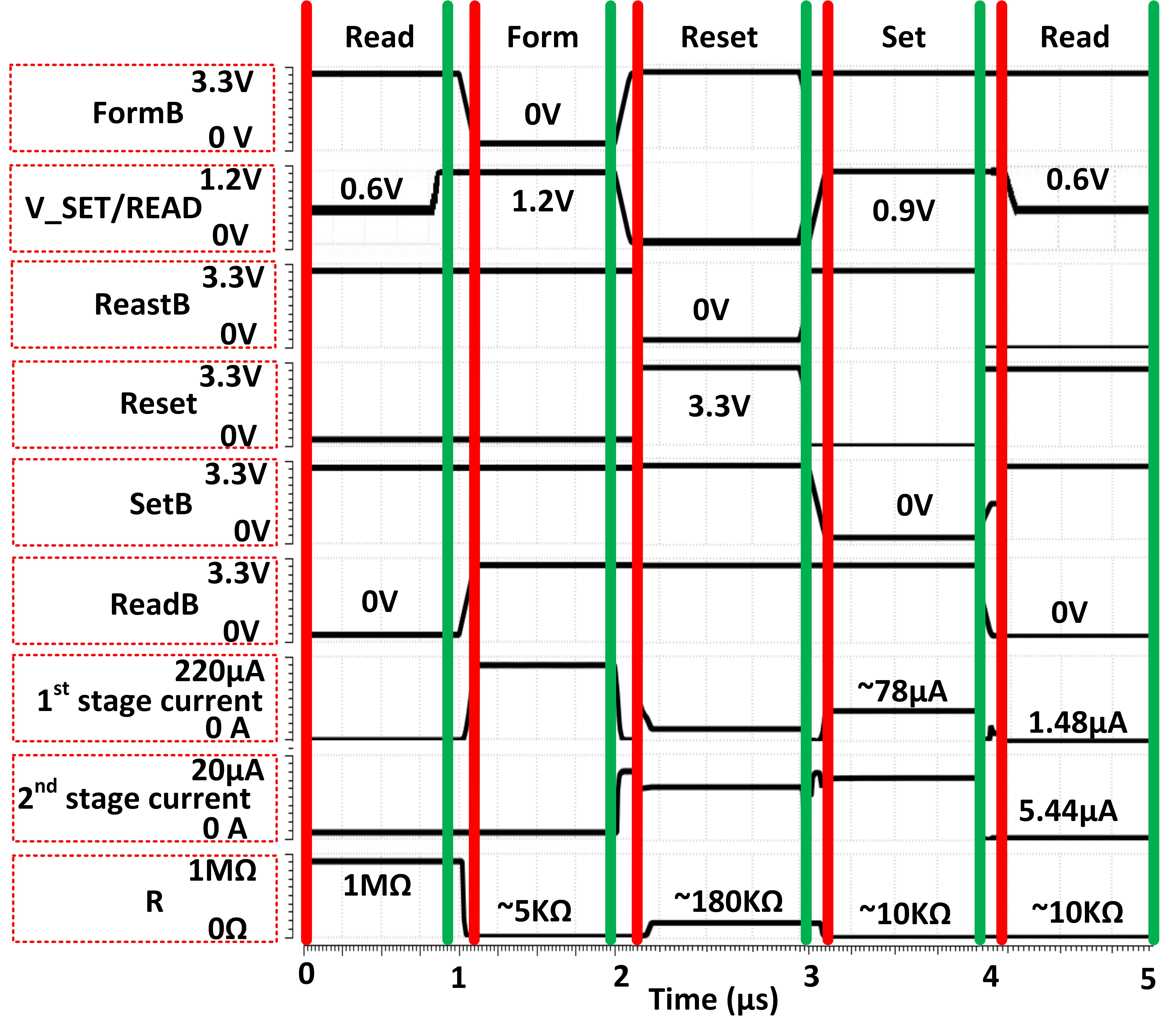}
            \caption{Timing diagram of our proposed synapse. \textit{SET} and \textit{READ} currents are labeled in the diagram.}
            \label{fig:timing}
        \end{figure}
        
\subsection{ Timing Diagram}
 A timing diagram of our proposed synapse is illustrated in Fig. \ref{fig:timing}. Each operation is managed by a pair of CMOS devices. At first, a \textit{READ} operation occurs before forming. During a \textit{READ} operation, V\_SET\slash READ and ReadB are set at \SI{0.6}{\volt} and \SI{0}{\volt} respectively. Initially, the value of the memristor is about \SI{1}{\mega\ohm}. Hence a \textit{FORM} operation is initiated by enabling \textit{FormB} and V\_SET\slash READ signal. As a result, the memristor value drops to \SI{5}{\kilo\ohm}. After that, a \textit{RESET} operation is applied by \textit{ResetB} and \textit{Reset} signals. The value of the memristor hikes to $\sim$\SI{180}{\kilo\ohm}. Then a SET operation is enabled with SET signals. At \SI{0.9}{\volt} for V\_SET\slash READ, the memristor value drops to $\sim$\SI{10}{\kilo\ohm}. Finally, a \textit{READ} operation is observed to verify the \textit{SET} value in the memristor.  
\subsection{ Simulation Results: SET Variation}

\begin{figure}[]
            \centering
            \includegraphics[width=3.4in]{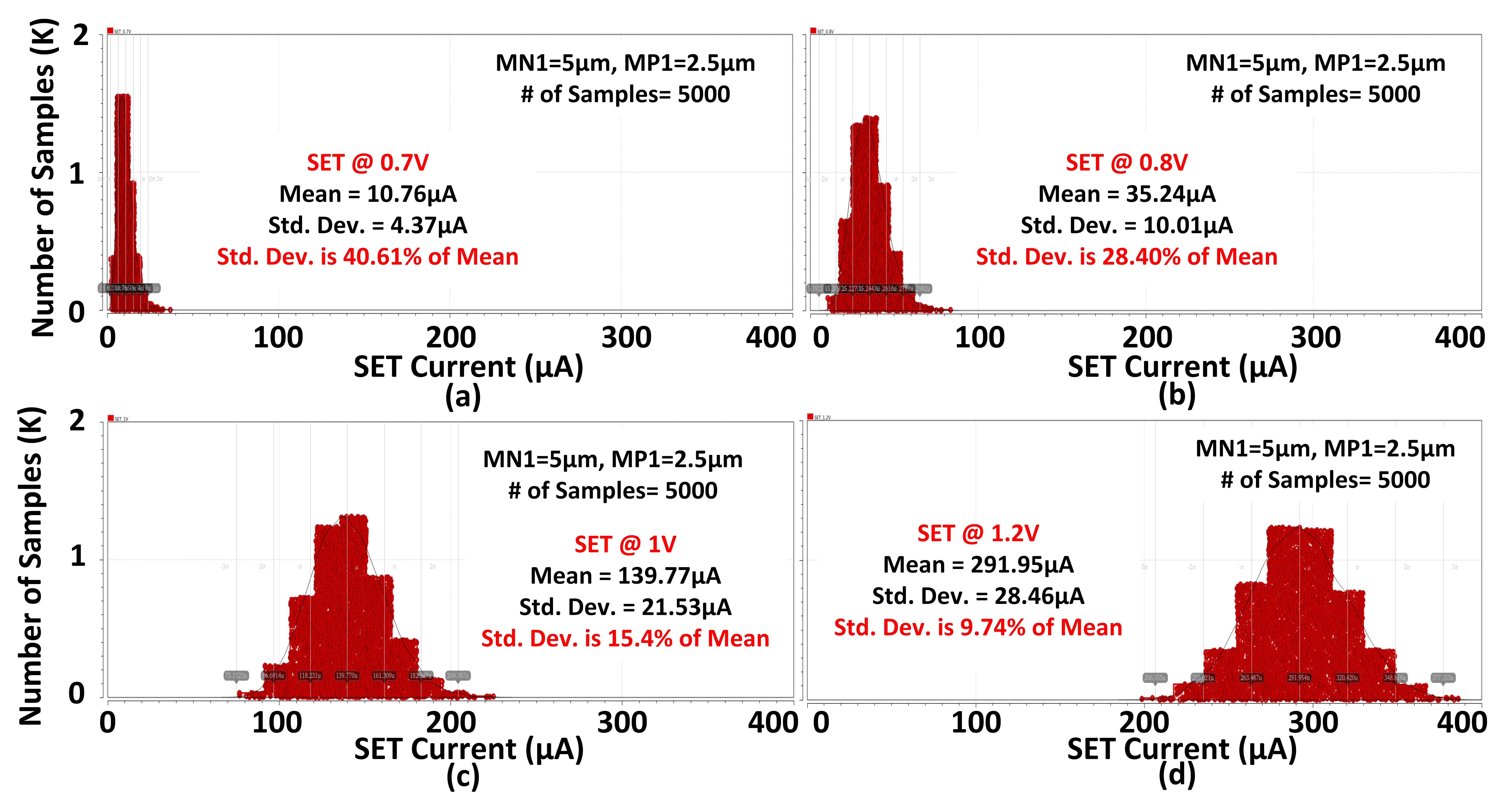}
            \caption{\textit{SET} current variation at different programming or \textit{SET} voltages. (a) and (b) show the \textit{SET} variation at \SI{0.7}{\volt} and \SI{0.8}{\volt}, where the ratio of std. dev. and mean reduced from 40\% to 28\%. (c) and (d) show, the ratio is reduced from 15.4\% to 9.74\%. At LRS or higher \textit{SET} voltage, our proposed design shows stable programming characteristics.}
            \label{fig:SET_Vriation}
        \end{figure}

At first the \textit{SET} operation is considered, where different \textit{SET} or programming voltages are applied to write different memristor levels. The main reason of this analysis is to define the process variation sensitivity of the design for the \textit{SET} operation. Fig. \ref{fig:SET_Vriation} shows the \textit{SET} variation when the gate voltage of $M_{N1}$ is \SI{0.7}{\volt}, \SI{0.8}{\volt}, \SI{1.0}{\volt}, and \SI{1.2}{\volt}. At lower \textit{SET} voltages, the memristor will be set at higher resistance levels and vice versa. The width of $M_{N1}$ and $M_{P1}$ are set at \SI{5}{\micro\meter} and \SI{2.5}{\micro\meter} respectively. The length for both devices remain at \SI{0.5}{\micro\meter}. Fig. \ref{fig:SET_Vriation} (a), exhibits the process variation sensitivity of the \textit{SET} current with \SI{0.7}{\volt}. The mean and std. dev. are \SI{10.76}{\micro\ampere} and \SI{4.37}{\micro\ampere} accordingly. Here, the std. dev. is about 40\% of the mean \textit{SET} current. Fig. \ref{fig:SET_Vriation} (b) and (c) show the \textit{SET} variability at \SI{0.8}{\volt} and \SI{1}{\volt}, where the ratio of std. dev. and mean is reduced from 28\% to 15.4\%. Finally, Monte Carlo analysis is performed for the \textit{SET} operation at \SI{1.2}{\volt}. The ratio at \SI{1.2}{\volt} is 9.74\%, which is much lower than at \SI{0.7}{\volt}. As a result, increasing the voltage during \textit{SET} will reduce the process variation sensitivity.              

\subsection{Simulation Results: READ Variation}

\begin{figure}[t]
            \centering
            \includegraphics[width=3.4in]{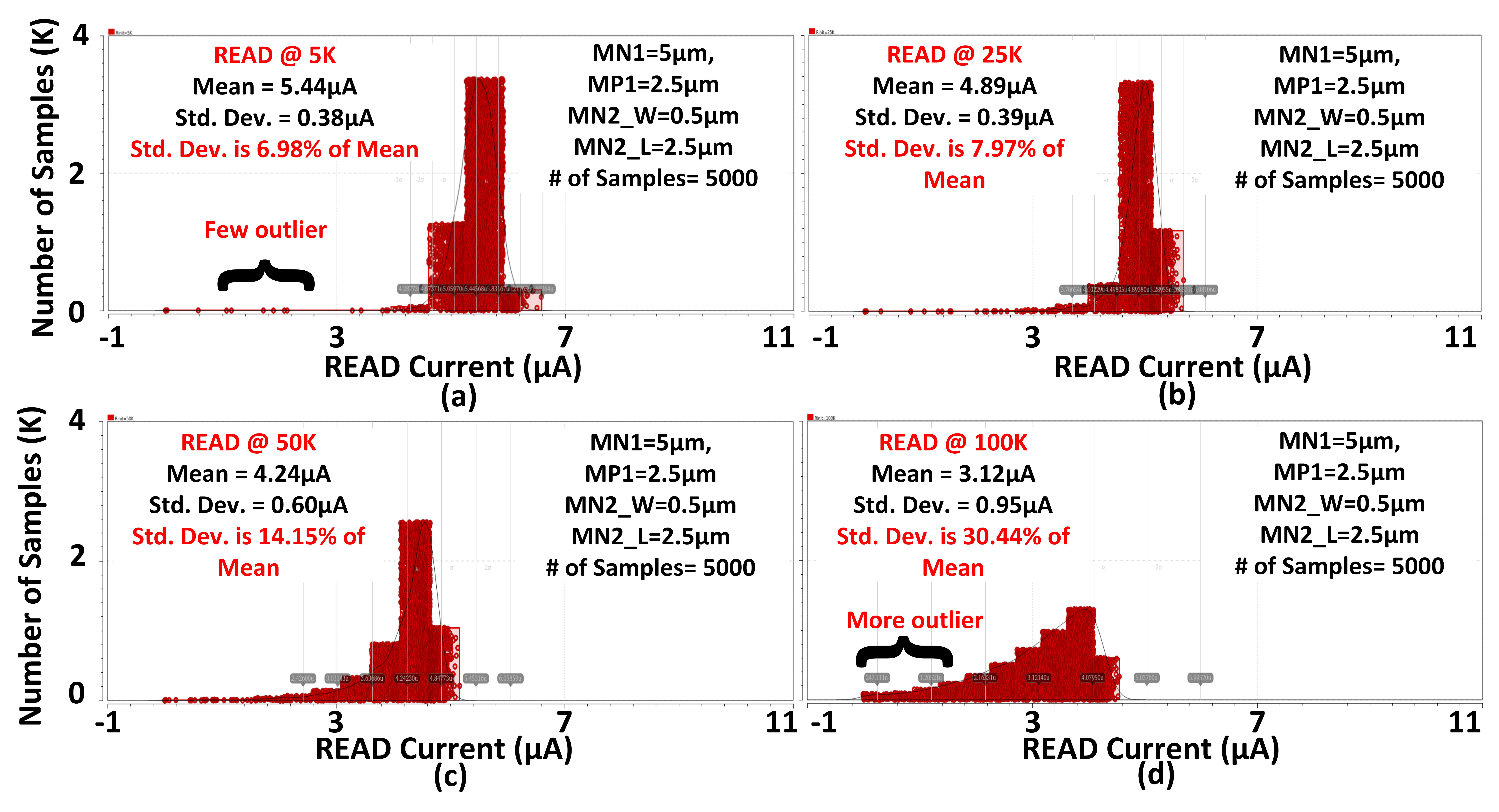}
            \caption{\textit{READ} current CMOS process variation effects at different resistance levels. (a) shows the \textit{READ} current variation at \SI{5}{\kilo\ohm}. The std. dev. and the mean current ratio is about 6.98\%. (b) illustrates the ratio is 7.97\% at \SI{25}{\kilo\ohm}, which is a bit higher than the last test case. (c) and (d) show the \textit{READ} current variability at \SI{50}{\kilo\ohm} and \SI{100}{\kilo\ohm} respectively. (d) shows the ratio is about 30.44\%, which is the maximum among all the test cases.}
            \label{fig:READ_Vriation}
        \end{figure}

The \textit{READ} operation of the memristor is highly susceptible to process variation. Fig. \ref{fig:READ_Vriation} shows the \textit{READ} variation at different memristor levels. Here the \textit{READ} voltage is \SI{0.6}{\volt} and the width of $M_{N1}$ and $M_{P1}$ are fixed at \SI{5}{\micro\meter} and \SI{2.5}{\micro\meter} respectively. The length for both $MOSFETs$ is \SI{0.5}{\micro\meter}. In addition, the width and length of $M_{N2}$ are fixed at \SI{0.5}{\micro\meter} and \SI{2.5}{\micro\meter} accordingly. 5000 samples are taken under consideration for MC simulation. Fig. \ref{fig:READ_Vriation} (a) shows the \textit{READ} variation at \SI{5}{\kilo\ohm}. The mean and std. dev. are \SI{5.44}{\micro\ampere} and \SI{0.38}{\micro\ampere} respectively. The ratio of the std. dev. and mean current is about 6.98\%. Fewer outliers are observed in this test case. Fig. \ref{fig:READ_Vriation} (b) and (c) show the \textit{READ} current variation at \SI{25}{\kilo\ohm} and \SI{50}{\kilo\ohm} respectively. In these two cases, the ratio increased from 7.97\% to 14.15\%. At \SI{100}{\kilo\ohm}, this ratio increased to 30.44\%. These results clearly indicate at LRS the effect of CMOS process variation is significantly reduced compared to HRS. In the next subsection, a similar analysis will be performed with measured data from fabricated devices.        

\subsection{ Measured Results}
\begin{figure}[t]
            \centering
            \includegraphics[width=3.4in]{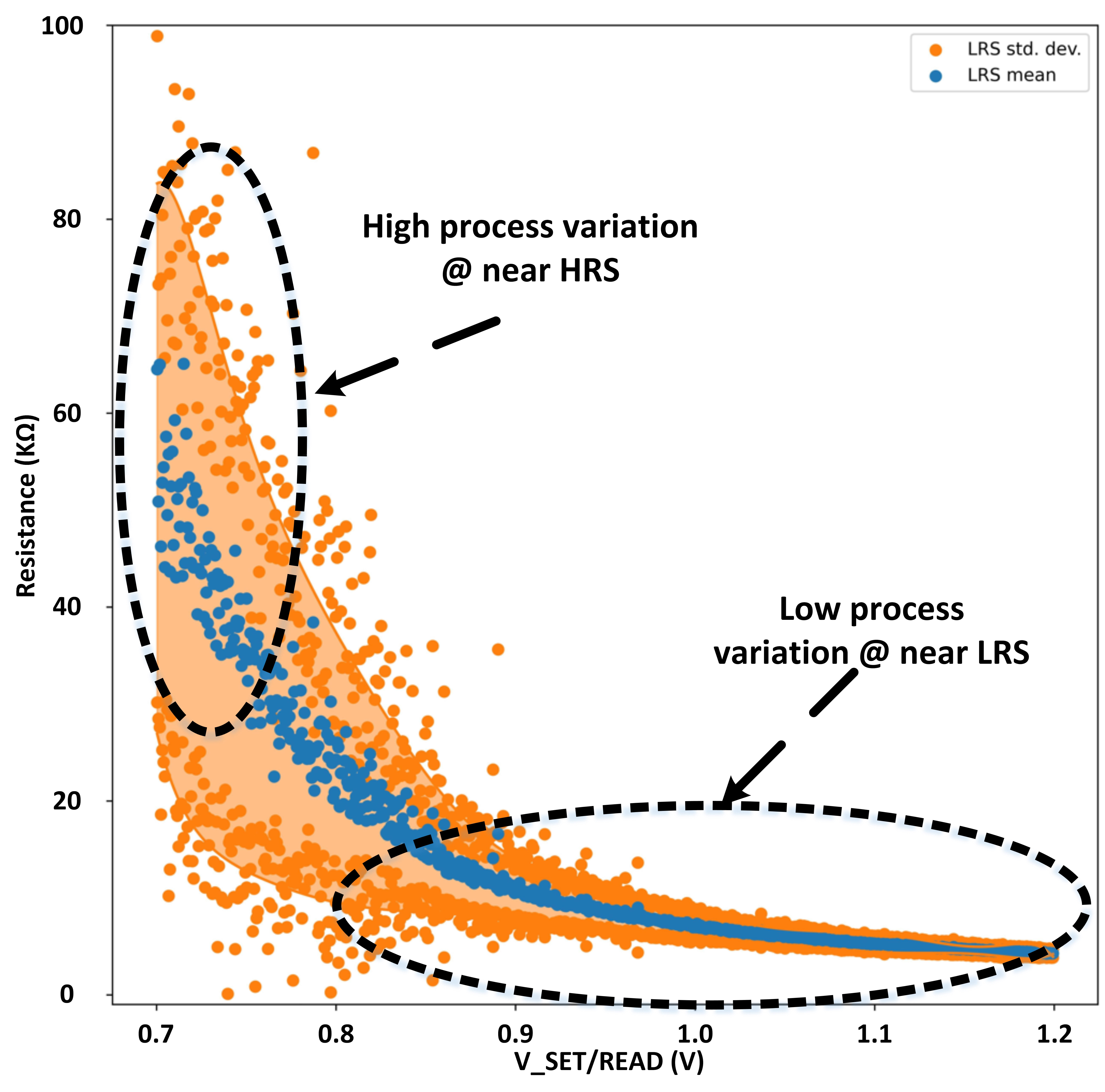}
            \caption{Mean and std. dev. of LRS resistance measured across 10 devices as \textit{SET} voltage at the gate of $M_{N1}$ is varied during a \textit{SET} operation. The mean resistance in blue is shown with a highlighted area above and below representing the std. dev. According to the measured data, the stochastic memristive behavior is less present at LRS compared to HRS. At LRS the process variation about 9\% whereas at HRS the process variation is about 31\%.} 
            \label{fig:analog_programming_LRS_multi_device}
        \end{figure}

In this section, measured results are shown. Unlike ealier Monte Carlo simulations, these measurements are also effected by the inherent stochastic behavior of the memristor. Memristors and synapses are fabricated based on \ce{HfO2} material in \SI{65}{\nano\meter} CMOS process. A source meter is connected to the probe station to perform DC analysis during \textit{FORM}, \textit{RESET}, \textit{SET}, and \textit{READ} operations. This process uses Python to automate application of voltage and current in addition to the collection of data. Commands are sent over TCP/IP to the sourcemeter from the host machine running the Python code. Measurements show high device to device variation as well as singular device variation in stored resistance for HRS. This variation also is shown in the higher resistance range of LRS. However, in the lower LRS the relation between the gate voltage and the resulting resistance becomes more predictable with a lower std. dev. and linear relationship. It can be seen in Fig. \ref{fig:analog_programming_LRS_multi_device} that in lower \textit{SET} voltages the stored resistance value has a high std. dev. which causes unpredictable behavior. However, at approximately \SI{20}{\kilo\ohm} or \SI{800}{\milli\volt} the standard deviation begins to drop sharply.

\section{Proposed Design for Applications}

\subsection{Synapse as a 4-bit Memory}
Table \ref{tab:4-bit} demonstrates the severe performance impact the lack of proper sizing can have on a memristive synapse. Our proposed synapse can be utilized as a 4-bit memory element. This synapse can be programmed  from \SI{5}{\kilo\ohm} to \SI{20}{\kilo\ohm} with a \textit{SET} voltage from \SI{0.8}{\volt} to \SI{1.2}{\volt} respectively. According to TABLE \ref{tab:4-bit}, \SI{5}{\kilo\ohm} is mapped to "0000", \SI{6}{\kilo\ohm} is mapped to "0001" and so on. There are 6 possible sizing combinations for \textit{READ} operation.
According to TABLE II, the $1^{st}$ stage “0000” is considered as the base current. Each case demonstrates the effects of the transistor sizing on the $2^{nd}$ stage / READ current.  Hence, if the $2^{nd}$ state “0001” shows \SI{20}{\nano\ampere} more than the $1^{st}$ stage then the $2^{nd}$ state will be also sensible, and so on. In case 2, the $1^{st}$ and $2^{nd}$ states have at least \SI{20}{\nano\ampere} READ current differences between the states. After that, all the READ current levels show the same current as the $2^{nd}$ state. Due to that, only two states are recognizable with \SI{20}{\nano\ampere} current differences. However, case 6  shows at least a \SI{20}{\nano\ampere} current difference between the consecutive resistance states. Thus all 16 levels are sensible. 
In column-3, 12.5\% states are readable with \SI{20}{\nano\ampere} resolution-based ADC or neuron for the neuromorphic system. This sizing shows a minimal design area but should be avoided due to severe precision limitations. The next test case shows with the width of $M_{P1}$ increased to \SI{2.5}{\micro\meter} still only 12.5\% of states can be read via the same ADC circuit. If the width of $M_{N1}$ is increased to \SI{5}{\micro\meter} and all other devices are of minimal size, only 18.75\% of states can be read. Hence, if the widths of $M_{N1}$ and $M_{P1}$ are \SI{5}{\micro\meter} and \SI{2.5}{\micro\meter} and $M_{N2}$ is \SI{0.5}{\micro\meter}, then only 43.75\% states are readable. Later, the width of $M_{N2}$ is increased to \SI{2.5}{\micro\meter} and the readable states are 12.5\%. Finally, using the sized shown in \ref{tab:sizing}, all states are readable with a \SI{20}{\nano\ampere} resolution sense amplifier and ADC. Our main focus of this work is to design a reliable synapse for a neuromorphic core, where the synaptic current will be received by a CMOS neuron. 


\subsection{Application Tests with the TENNLAB Neuromorphic Framework}
The TENNLAB exploratory neuromorphic computing framework~\cite{psb:18:ten} facilitates the development of spiking neural networks (SNNs) on problems in many application domains including the the control theory domain and classification domain. SNNs are neuromorphic computing based networks which can be mapped onto neuromorphic hardware. In this software SNNs are generated using the evolutionary optimization for neuromorphic systems (EONS)~\cite{smp:20:eons}. EONS is a genetic algorithm that evolves populations of networks by applying genetic operators over the course of many epochs.

\subsection{Mapping Circuit Based Weights into the Framework}
In order to represent the effect of the different circuit cases, the ideal weight precision needed to be reduced in a way which makes sense for a synapse in neuromorphic hardware. To do this, design cases with low-bit representations, which represent small changes in current between each bit value, were mapped into small values in the network. For instance, in case 2 since there are only two differentiable \textit{READ} states for the synapse, the synapse can only represent two weight values. So, if the synapse in the ideal network contained a weight value of 15, this value would be converted to 1. 
The last row of TABLE \ref{tab:4-bit} shows the possible synaptic weight values for every test case. The networks were trained with all possible weight values, ie. [0,15] using EONS [25]. In order to evaluate the effects of a limited set of synaptic weight values, the networks with ideal weight values were converted into networks containing only weights from the limited set (test case) using TABLE \ref{tab:4-bit}.
After converting all of the weights in the ideal network into weights which represent the design cases, the task was then run again to obtain a testing metric. This same process was carried out for both classification and control applications. 
For our experiment, only positive weight values are considered.

Here, elements such as parasitics are not considered for neuromorphic framework verification. Only the synapse behavior and parameters like \textit{READ} current resolution and power consumption are considered as a look-up table information to measure or evaluate the performance of SNNs by EONS. This design can be utilized in an array design by summing the second stage current of multiple synapse circuits.

\begin{figure}[t]
            \centering
            \includegraphics[width=3.4in]{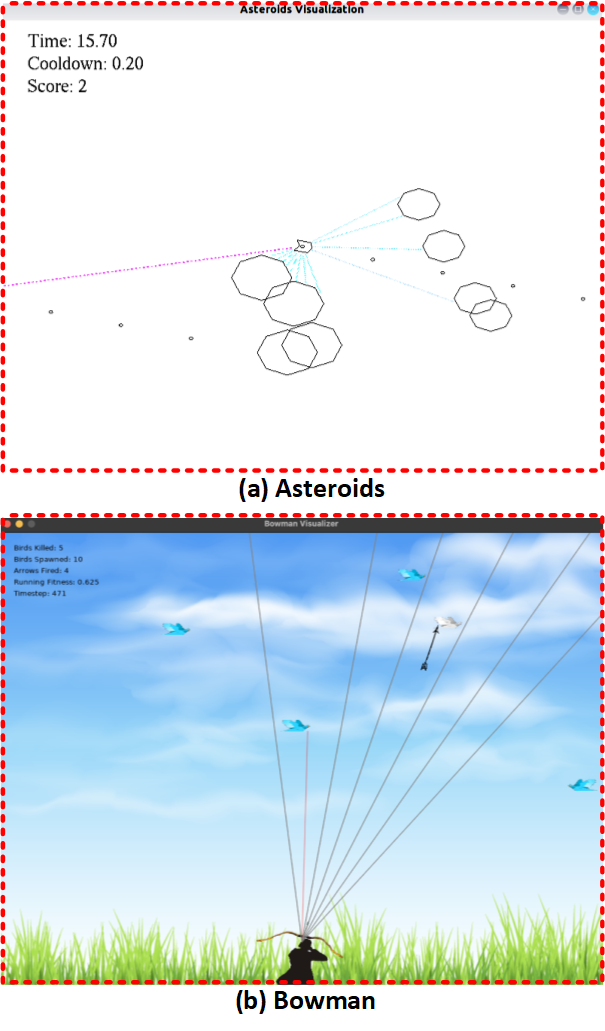}
            \caption{Screenshot of control applications. (a) illustrates the control application called Asteroids. The fitness of Asteroids depends on how long the agent can survive in the environment. (b) shows another control application called Bowman. The fitness function of Bowman depends on how many birds it can target with a fewer number of arrows. Asteroids is more complex than Bowman.}
            \label{fig:control_app}
        \end{figure}

\subsection{Synapse for Control Applications}
Control theory applications are popular evaluation tasks in neuromorphic computing. Two control applications that the framework natively supports are shown in Fig. \ref{fig:control_app} (a) Asteroids and (b) Bowman. 

Asteroids is based off of the classical Atari asteroids game. It is a complex game in which asteroids of different sizes are randomly generated, and the player needs to shoot and avoid the asteroids to remain alive and maximize the player's score. Its score (or fitness) is based on the agent's ability to both stay alive and shoot asteroids, and it is presented mathematically in an earlier work~\cite{prs:19:tsl}. The agent's training score is based on the score achieved using scenarios observed during the development of the network. Its testing score is based on scenarios not seen during the development of the network. The agent's training and testing fitnesses have a maximum value of 1.0. Due to the complexity of asteroids trained agents tend to have lower fitnesses in the 0.2-0.4 range~\cite{r23}. 

\begin{table*}[t]

\centering
    \caption{Synapse as a 4-bit memory element}
  
{
\begin{tabular}{|cc|c|c|c|c|c|c|c|}
\hline
\multicolumn{1}{|c|}{N/A} &
  {N/A} &
  {Case 1} &
  {Case 2} &
  {Case 3} &
  {Case 4} &
  {Case 5} &
  {Case 6} \\ \hline
\multicolumn{1}{|c|}{\begin{tabular}[c]{@{}l@{}}Resistance \\ ( K  $\Omega$)\end{tabular}} &
  {\begin{tabular}[c]{@{}l@{}}Corresponding\\  Bit\end{tabular}} &
  {\begin{tabular}[c]{@{}l@{}}$M_{N1}$=1µm, \\ $M_{P1}$=0.5µm, \\ $M_{N2_W}$=0.5µm,   \\ $M_{N2_L}$=0.5µm\end{tabular}} &
  {\begin{tabular}[c]{@{}l@{}}$M_{N1}$=1µm, \\ $M_{P1}$=2.5µm,\\ $M_{N2_W}$=0.5µm,\\ $M_{N2_L}$=0.5µm\end{tabular}} &
  {\begin{tabular}[c]{@{}l@{}}$M_{N1}$=5µm,\\  $M_{P1}$=0.5µm, \\ $M_{N2_W}$=0.5µm,\\    $M_{N2_L}$=0.5µm\end{tabular}} &
  {\begin{tabular}[c]{@{}l@{}}$M_{N1}$=5µm, \\ $M_{P1}$=2.5µm, \\ $M_{N2_W}$=0.5µm,\\    $M_{N2_L}$=0.5µm\end{tabular}} &
  {\begin{tabular}[c]{@{}l@{}}$M_{N1}$=5µm,\\  $M_{P1}$=2.5µm,\\    $M_{N2_W}$=2.5µm, \\ $M_{N2_L}$=0.5µm\end{tabular}} &
  {\begin{tabular}[c]{@{}l@{}}$M_{N1}$=5µm, \\ $M_{P1}$=2.5µm,\\    $M_{N2_W}$=0.5µm,\\  $M_{N2_L}$=2.5µm\end{tabular}} \\ \hline
\multicolumn{1}{|c|} {\color[HTML]{000000} {5}} &
  {\color[HTML]{000000} {0000}} &
  {\color[HTML]{000000} {\checkmark}} &
  {\color[HTML]{000000} {\checkmark}} &
  {\color[HTML]{000000} {\checkmark}} &
  {\color[HTML]{000000} {\checkmark}} &
  {\color[HTML]{000000} {\checkmark}} &
  {\color[HTML]{000000} {\checkmark}} \\ \hline
\multicolumn{1}{|c|} {\color[HTML]{000000} {6}} &
  {\color[HTML]{000000} {0001}} &
  {\color[HTML]{000000} {x}} &
  {\color[HTML]{000000} {\checkmark}} &
  {\color[HTML]{000000} {x}} &
  {\color[HTML]{000000} {x}} &
  {\color[HTML]{000000} {x}} &
  {\color[HTML]{000000} {\checkmark}} \\ \hline
\multicolumn{1}{|c|} {\color[HTML]{000000} {7}} &
  {\color[HTML]{000000} {0010}} &
  {\color[HTML]{000000} {\checkmark}} &
  {\color[HTML]{000000} {x}} &
  {\color[HTML]{000000} {x}} &
  {\color[HTML]{000000} {x}} &
  {\color[HTML]{000000} {x}} &
  {\color[HTML]{000000} {\checkmark}} \\ \hline
\multicolumn{1}{|c|} {\color[HTML]{000000} {8}} &
  {\color[HTML]{000000} {0011}} &
  {\color[HTML]{000000} {x}} &
  {\color[HTML]{000000} {x}} &
  {\color[HTML]{000000} {x}} &
  {\color[HTML]{000000} {\checkmark}} &
  {\color[HTML]{000000} {x}} &
  {\color[HTML]{000000} {\checkmark}} \\ \hline
\multicolumn{1}{|c|} {\color[HTML]{000000} {9}} &
  {\color[HTML]{000000} {0100}} &
  {\color[HTML]{000000} {x}} &
  {\color[HTML]{000000} {x}} &
  {\color[HTML]{000000} {x}} &
  {\color[HTML]{000000} {x}} &
  {\color[HTML]{000000} {x}} &
  {\color[HTML]{000000} {\checkmark}} \\ \hline
\multicolumn{1}{|c|} {\color[HTML]{000000} {10}} &
  {\color[HTML]{000000} {0101}} &
  {\color[HTML]{000000} {x}} &
  {\color[HTML]{000000} {x}} &
  {\color[HTML]{000000} {x}} &
  {\color[HTML]{000000} {\checkmark}} &
  {\color[HTML]{000000} {x}} &
  {\color[HTML]{000000} {\checkmark}} \\ \hline
\multicolumn{1}{|c|} {\color[HTML]{000000} {11}} &
  {\color[HTML]{000000} {0110}} &
  {\color[HTML]{000000} {x}} &
  {\color[HTML]{000000} {x}} &
  {\color[HTML]{000000} {x}} &
  {\color[HTML]{000000} {x}} &
  {\color[HTML]{000000} {x}} &
  {\color[HTML]{000000} {\checkmark}} \\ \hline
\multicolumn{1}{|c|} {\color[HTML]{000000} {12}} &
  {\color[HTML]{000000} {1011}} &
  {\color[HTML]{000000} {x}} &
  {\color[HTML]{000000} {x}} &
  {\color[HTML]{000000} {\checkmark}} &
  {\color[HTML]{000000} {x}} &
  {\color[HTML]{000000} {x}} &
  {\color[HTML]{000000} {\checkmark}} \\ \hline
\multicolumn{1}{|c|} {\color[HTML]{000000} {13}} &
  {\color[HTML]{000000} {1000}} &
  {\color[HTML]{000000} {x}} &
  {\color[HTML]{000000} {x}} &
  {\color[HTML]{000000} {x}} &
  {\color[HTML]{000000} {\checkmark}} &
  {\color[HTML]{000000} {\checkmark}} &
  {\color[HTML]{000000} {\checkmark}} \\ \hline
\multicolumn{1}{|c|} {\color[HTML]{000000} {14}} &
  {\color[HTML]{000000} {1001}} &
  {\color[HTML]{000000} {x}} &
  {\color[HTML]{000000} {x}} &
  {\color[HTML]{000000} {x}} &
  {\color[HTML]{000000} {x}} &
  {\color[HTML]{000000} {x}} &
  {\color[HTML]{000000} {\checkmark}} \\ \hline
\multicolumn{1}{|c|} {\color[HTML]{000000} {15}} &
  {\color[HTML]{000000} {1010}} &
  {\color[HTML]{000000} {x}} &
  {\color[HTML]{000000} {x}} &
  {\color[HTML]{000000} {x}} &
  {\color[HTML]{000000} {\checkmark}} &
  {\color[HTML]{000000} {x}} &
  {\color[HTML]{000000} {\checkmark}} \\ \hline
\multicolumn{1}{|c|} {\color[HTML]{000000} {16}} &
  {\color[HTML]{000000} {1011}} &
  {\color[HTML]{000000} {x}} &
  {\color[HTML]{000000} {x}} &
  {\color[HTML]{000000} {x}} &
  {\color[HTML]{000000} {x}} &
  {\color[HTML]{000000} {x}} &
  {\color[HTML]{000000} {\checkmark}} \\ \hline
\multicolumn{1}{|c|} {\color[HTML]{000000} {17}} &
  {\color[HTML]{000000} {1100}} &
  {\color[HTML]{000000} {x}} &
  {\color[HTML]{000000} {x}} &
  {\color[HTML]{000000} {x}} &
  {\color[HTML]{000000} {x}} &
  {\color[HTML]{000000} {x}} &
  {\color[HTML]{000000} {\checkmark}} \\ \hline
\multicolumn{1}{|c|} {\color[HTML]{000000} {18}} &
  {\color[HTML]{000000} {1101}} &
  {\color[HTML]{000000} {x}} &
  {\color[HTML]{000000} {x}} &
  {\color[HTML]{000000} {x}} &
  {\color[HTML]{000000} {\checkmark}} &
  {\color[HTML]{000000} {x}} &
  {\color[HTML]{000000} {\checkmark}} \\ \hline
\multicolumn{1}{|c|} {\color[HTML]{000000} {19}} &
  {\color[HTML]{000000} {1110}} &
  {\color[HTML]{000000} {x}} &
  {\color[HTML]{000000} {x}} &
  {\color[HTML]{000000} {\checkmark}} &
  {\color[HTML]{000000} {x}} &
  {\color[HTML]{000000} {x}} &
  {\color[HTML]{000000} {\checkmark}} \\ \hline
\multicolumn{1}{|c|} {\color[HTML]{000000} {20}} &
  {\color[HTML]{000000} {1111}} &
  {\color[HTML]{000000} {x}} &
  {\color[HTML]{000000} {x}} &
  {\color[HTML]{000000} {x}} &
  {\color[HTML]{000000} {\checkmark}} &
  {\color[HTML]{000000} {x}} &
  {\color[HTML]{000000} {\checkmark}} \\ \hline
 \multicolumn{2}{|c|} {\color[HTML]{000000} \textbf{Readability}} &
  {\color[HTML]{000000} {12.5\%}} &
  {\color[HTML]{000000} {12.5\%}} &
  {\color[HTML]{000000} {18.75\%}} &
  {\color[HTML]{000000} {43.75\%}} &
  {\color[HTML]{000000} {12.5\%}} &
  {\color[HTML]{000000} {100\%}} \\ \hline 

  \multicolumn{2}{|c|} {\color[HTML]{000000} \textbf{Bit precision}} &
  {\color[HTML]{000000} {1}} &
  {\color[HTML]{000000} {1}} &
  {\color[HTML]{000000} {$<2$}} &
  {\color[HTML]{000000} {$<3$}} &
  {\color[HTML]{000000} {1}} &
  {\color[HTML]{000000} {4}} \\ \hline
  
  \multicolumn{2}{|c|} {\color[HTML]{000000} \textbf{Synaptic Weight Range}} &
  {\color[HTML]{000000} {[0,1]}} &
  {\color[HTML]{000000} {[0,1]}} &
  {\color[HTML]{000000} {[0,2]}} &
  {\color[HTML]{000000} {[0,6]}} &
  {\color[HTML]{000000} {[0,1]}} &
  {\color[HTML]{000000} {[0,15]}} \\ \hline
  
\end{tabular}%
}
\label{tab:4-bit}

\end{table*}

Bowman is based off of the ``Bowman2" minigame featured on crazygames.com. It features a stationary bowman at the bottom-center of the game window whose objective is to shoot birds that randomly spawn and fly overhead. Its fitness metric is also defined in~\cite{prs:19:tsl}, but unlike asteroids, there is no death mechanic, so the average training fitnesses tend to be in the 0.6-0.9 range. Both applications use simulated LIDAR to provide inputs to the agent.

For this work, one hundred networks for each Bowman and Asteroids were trained. The average results of these training experiments are displayed in TABLE \ref{tab:control} in the ``Default" row. The testing results for each mapping of the networks' synapses to different bit precision cases are also shown. The table shows that as the bit precision of the synaptic mapping decreases, so also does the application's testing performance. TABLE \ref{tab:control} shows test fitness is improved by about 53\% for \textit{Asteroids} when comparing to Case 6 over Case 1. Moreover, about 82\%  test fitness is improved for \textit{Bowman} compared to Cases 1 and 6. More bit precision shows better fitness scores for both control applications.

\subsection{Synapse for Classifications}
Classification tasks are evaluation staples in machine learning and can be achieved using synapse arrays. Following the same training methodology as was done with the control tasks, we use the TENNLAB software framework to learn spiking neural networks on two popular toy datasets: breast cancer~\cite{asuncion2007uci} and wine~\cite{asuncion2007uci}. TABLE \ref{tab:classifications} shows the results of the training runs. As with the Bowman and Asteroids tests, classification on the wine dataset followed a similar trend: mapping the network weights to lower bit precision values resulted in decreasing performance with case 6 matching the full bit precision performance. An interesting outlier for case 4 of the breast cancer classification experiment can be found. On the breast cancer classification task, both case 6 and the lower precision case 4 were able to match the accuracy of the default case. This implies that for some classification datasets (and even other control tasks), it might be possible to reduce the bit precision of the synapses that compose a network without a loss in performance. TABLE \ref{tab:classifications} shows, about 2.7x testing accuracy improvement with case 6 compared to case 1 for the Wine dataset. 

\begin{table}[]
\centering
    \caption{Design impact on control applications}
\begin{tabular}{|c|cc|cc|}
\hline
Dataset             & \multicolumn{2}{c|}{Asteroids } & \multicolumn{2}{c|}{Bowman } \\ \hline
State                         & \multicolumn{1}{c|}{Training} & Testing               & \multicolumn{1}{c|}{Training} & Testing               \\ \hline
\multicolumn{1}{|l|}{Default} & \multicolumn{1}{l|}{0.300}         & \multicolumn{1}{l|}{0.199} & \multicolumn{1}{l|}{0.686}         & \multicolumn{1}{l|}{0.473} \\ \hline
Fitness of Case 1  & \multicolumn{1}{c|}{--}      & {0.130}      & \multicolumn{1}{c|}{--}  & {0.260} \\ \hline
Fitness of Case 2  & \multicolumn{1}{c|}{--}      &  {0.130}     & \multicolumn{1}{c|}{--}  & {0.260} \\ \hline
Fitness of  Case 3 & \multicolumn{1}{c|}{--}      &  {0.168}     & \multicolumn{1}{c|}{--}  & {0.283} \\ \hline
Fitness of  Case 4 & \multicolumn{1}{c|}{--}      &  {0.187}     & \multicolumn{1}{c|}{--}  & {0.433} \\ \hline
Fitness of  Case 5 & \multicolumn{1}{c|}{--}      &  {0.130}     & \multicolumn{1}{c|}{--}  & {0.260} \\ \hline
Fitness of  Case 6 & \multicolumn{1}{c|}{--}      &  {0.199}     & \multicolumn{1}{c|}{--}  & {0.473} \\ \hline
\end{tabular}%
\label{tab:control}
\end{table}

\begin{table}[]
\centering
    \caption{Design impact on Classifications }
\begin{tabular}{|c|cc|cc|}
\hline
Dataset & \multicolumn{2}{c|}{Breast Cancer} & \multicolumn{2}{c|}{Wine} \\ \hline
State & \multicolumn{1}{c|}{Training} & Testing & \multicolumn{1}{c|}{Training} & Testing \\ \hline
Default             & \multicolumn{1}{l|}{0.956}     & {0.862} & \multicolumn{1}{l|}{0.897}         & {0.823} \\ \hline
Accuracy of Case 1  & \multicolumn{1}{c|}{--}      & {0.353} & \multicolumn{1}{c|}{--}  & {0.301} \\ \hline
Accuracy of Case 2  & \multicolumn{1}{c|}{--}      & {0.353} & \multicolumn{1}{c|}{--}  & {0.301} \\ \hline
Accuracy of  Case 3 & \multicolumn{1}{c|}{--}      & {0.537} & \multicolumn{1}{c|}{--}  & {0.378} \\ \hline
Accuracy of  Case 4 & \multicolumn{1}{c|}{--}      & {0.862} & \multicolumn{1}{c|}{--}  & {0.490} \\ \hline
Accuracy of  Case 5 & \multicolumn{1}{c|}{--}      & {0.353} & \multicolumn{1}{c|}{--}  & {0.301} \\ \hline
Accuracy of  Case 6 & \multicolumn{1}{c|}{--}      & {0.862} & \multicolumn{1}{c|}{--}  & {0.823} \\ \hline
\end{tabular}%
\label{tab:classifications}
\end{table}

\begin{table*}[]

\centering
    \caption{Comparison with Prior work per Synapse }
    
{
\begin{tabular}{|l|l|l|l|l|l|l|l|}
\hline
Ref.  & \cite{ref1} & \cite{ref2} & \cite{ref3} & \cite{ref4} & \cite{ref5} & \cite{ref6} & This Work                         \\ \hline
Process              & \SI{180}{\nano\meter} $CMOS$     &{x}     & {x}    &{x}     &{x}     & \SI{28}{\nano\meter} $CMOS$     & \SI{65}{\nano\meter} $CMOS$                       \\ \hline
Memristor Layer      & {x}    & \ce{TaN\slash HfO2\slash Pt}    & \begin{tabular}[c]{@{}l@{}} \ce{Cu\slash black \\ phosphorus\slash Au}  \end{tabular}    & \ce{Sn\slash HfO2\slash Pt}    & \ce{Nb\slash NiO\slash Nb}    & \ce{HZO}    & \ce{HfO2}
      \\ \hline
Sizing Info.         & {x}    &{\checkmark}     & {\checkmark}    & {\checkmark}    &{x}     &{x}     & {\checkmark}                      \\ \hline
Process Variation    & {x}    & {x}    & {\checkmark}    & {x}    & {\checkmark}    & {\checkmark}    & {\checkmark}                      \\ \hline
\begin{tabular}[c]{@{}l@{}} Programming \\   Region \end{tabular} & {x}    & LRS     & LRS     & {x}     & {x}    & {x}     & \begin{tabular}[c]{@{}l@{}} LRS \\ (5K$\Omega$   to 20K$\Omega$) \end{tabular} \\ \hline
Bit Precision        & up to 5-bit    & {x}     & {x}     & {x}    & {x}    & 2-bit    & 4-bit                                 \\ \hline
SET Power       & {x}   & \SI{1.6}{\milli\watt}    &  \SI{0.64}{\milli\watt}   & \SI{3.5}{\milli\watt}    & \SI{12.3}{\milli\watt}    & {x}    & \SI{0.46}{\milli\watt} @ \SI{1}{\volt}                                  \\ \hline
SET   Energy    &  {x}   & {x}     & {x}    & {x}    & {x}    & \SI{2.57}{\nano\joule} @ \SI{1}{\volt}    &  \SI{0.46}{\nano\joule} @ \SI{1}{\volt}   \\ \hline
READ   Power    &  \SI{27.43}{\micro\watt}     & {x}    & {x}     & {x}    & {x}    & {x}    & \begin{tabular}[c]{@{}l@{}} Avg. \SI{8.24}{\micro\watt} from \\ (5K$\Omega$   to 20K$\Omega$) \end{tabular}   \\ \hline
READ   Energy   & {x}    &  {x}   & {x}    & {x}    & {x}    & \SI{17.8}{\pico\joule}     & \begin{tabular}[c]{@{}l@{}} Avg. \SI{8.24}{\pico\joule} from \\ (5K$\Omega$   to 20K$\Omega$) \end{tabular}   \\ \hline
Measured Result      & {\checkmark}    & {\checkmark}      & {\checkmark}      & {\checkmark}      & {x}    & {\checkmark}     & {\checkmark}                               \\ \hline
\end{tabular}%
}

\label{tab:comparision}

\end{table*}

\section{Comparison With Prior Work}  

Synapses are utilized for different purposes such as a dot product engines \textit(DPEs) \cite{r21}, Spike timing dependent plasticity \textit (STDP) \cite{r22}, neuroprocessors \cite{r23}, and so on. Synapses can be constructed with different materials. Here, a brief comparison among different types of synapses is analyzed based on their sizing, programming region, bit precision, power, energy, and measured results. A CMOS-based synapse is present with up to 5-bit precision, which consumes  \SI{27.43}{\micro\watt} as \textit{READ} power in \cite{ref1}. \cite{ref1} shows no clear indication of sizing, process variation, and programming region. In \cite{ref2}, a \ce{HfO2} based synapse is proposed, which is programmed at LRS and consumes \SI{1.6}{\milli\watt} as \textit{SET} power. However, this design does not consider process variation and bit precision. Another research group presented a \ce{black phosphorus} based synapse with \SI{0.64}{\milli\watt} power consummation. Sizing, process variation, programming region, bit precision, and \textit{READ} power are not considered in \cite{ref3}.

In \cite{ref4}, authors illustrated a \ce{HfO2} based synapse which consumes \SI{3.5}{\milli\watt} as \textit{SET} power. There is no information about sizing, process variation analysis, bit precision, and \textit{READ} power. A \ce{NiO} based synapse presented, which consumes \SI{12.3}{\milli\watt} as \textit{SET} power in \cite{ref5}. This is higher than all other synapses compared here. A \ce{HZO} based synapse fabricated in \SI{28}{\nano\meter} process with 2-bit precision in \cite{ref6}. This design consumes \SI{2.57}{\nano\joule} and \SI{17.8}{\pico\joule} as \textit{SET} and \textit{READ} energy respectively. Finally, our proposed design is based on \ce{HfO2} and fabricated on \SI{65}{\nano\meter} CMOS process. Sizing of all the \textit{READ} and \textit{SET} devices are properly mentioned here with sizing effect. In addition, process variations are considered in simulation as well as measured results and analysis during use in neuromorphic applications. This synapse is suitable for programming in LRS with less process variation effects. \textit{SET} power is at least 28\% lower than all other designs. \textit{SET} energy is about 82\% lower when compared to the \ce{HZO} based synapse. Also, our design shows about 70\% \textit{READ} power savings compared to \cite{ref1}. \textit{READ} energy shows about 54\% savings with our proposed design compared to \cite{ref6}. 
Our synapse can be treated as a synaptic weight with a 4-bit resolution. With this resolution, it can be evaluated using a neuromorphic system like EONS. Our proposed synapse utilizes a CMOS neuron to accumulate the charge and fire event. The neuron consumes about \SI{12.5}{\pico\joule} energy for a fire event \cite{Gangotree}. It can be concluded, that our proposed design is reliable and low-power with 4-bit precision.  

Moreover, 3T1R consumes less \textit{READ} power compared to the 1T1R structure. Though 3T1R has more devices, the \textit{READ} current is significantly low for its current controlled operation. The READ current of 3T1R is more linear than the 1T1R design. 3T1R adds area overhead compared to the 1T1R structure \cite{TCAS}.

\section{Conclusions and future work}

Our analysis of the proposed memristive synapse focused on different sizing corners to produce a design which is optimized across many domains. This optimized design reduced the effects of process variation on \textit{SET} and \textit{READ} operations. In addition, to obtain low power design goals, different \textit{READ} and \textit{SET} voltages have been taken under consideration. Approximately 28\% \textit{SET} power improvements can be seen compared to the state-of-the-art. Moreover, about 70\% \textit{READ} power is optimized compared to prior works. To show the severity of the effects improper sizing can have, an application based experiment was performed. In this experiment, six test cases were chosen and analyzed using a neuromorphic application. Our synapse was utilized to explore control tasks with up to 82\% test fitness improvement and classification tasks with up to 2.7x testing accuracy improvement. This design shows 4-bit data precision with a \SI{20}{\nano\ampere} \textit{READ} current resolution. Other experiments can be performed outside of SNNs which will provide a better understanding of characteristics important to synapses. Future designs will focus on increasing the \textit{READ} current resolution in order to improve task performance. 


    \bibliographystyle{IEEEtran}
    \bibliography{bibliography}

\begin{thebibliography}{10}
\providecommand{\url}[1]{#1}
\csname url@samestyle\endcsname
\providecommand{\newblock}{\relax}
\providecommand{\bibinfo}[2]{#2}
\providecommand{\BIBentrySTDinterwordspacing}{\spaceskip=0pt\relax}
\providecommand{\BIBentryALTinterwordstretchfactor}{4}
\providecommand{\BIBentryALTinterwordspacing}{\spaceskip=\fontdimen2\font plus
\BIBentryALTinterwordstretchfactor\fontdimen3\font minus \fontdimen4\font\relax}
\providecommand{\BIBforeignlanguage}[2]{{%
\expandafter\ifx\csname l@#1\endcsname\relax
\typeout{** WARNING: IEEEtran.bst: No hyphenation pattern has been}%
\typeout{** loaded for the language `#1'. Using the pattern for}%
\typeout{** the default language instead.}%
\else
\language=\csname l@#1\endcsname
\fi
#2}}
\providecommand{\BIBdecl}{\relax}
\BIBdecl

\bibitem{r12}
Y.~Xu, H.~Das, Y.~Gong, and N.~Gong, ``On mathematical models of optimal video memory design,'' \emph{IEEE Transactions on Circuits and Systems for Video Technology}, vol.~30, no.~1, pp. 256--266, 2020.

\bibitem{r13}
J.~Edstrom, H.~Das, Y.~Xu, and N.~Gong, ``Memory optimization for energy-efficient differentially private deep learning,'' \emph{IEEE Transactions on Very Large Scale Integration (VLSI) Systems}, vol.~28, no.~2, pp. 307--316, 2020.

\bibitem{r14}
Y.~Xu, H.~Das, and N.~Gong, ``Application-aware quality-energy optimization: Mathematical models enabled simultaneous quality and energy-sensitive optimal memory design,'' \emph{IEEE Transactions on Sustainable Computing}, vol.~6, no.~4, pp. 559--571, 2021.

\bibitem{r1}
L.~Chua, ``Memristor-the missing circuit element,'' \emph{IEEE Transactions on Circuit Theory}, vol.~18, no.~5, pp. 507--519, 1971.

\bibitem{r15}
S.~Gupta, M.~Imani, and T.~Rosing, ``Felix: Fast and energy-efficient logic in memory,'' in \emph{2018 IEEE/ACM International Conference on Computer-Aided Design (ICCAD)}, 2018, pp. 1--7.

\bibitem{r16}
R.~Gharpinde, P.~L. Thangkhiew, K.~Datta, and I.~Sengupta, ``A scalable in-memory logic synthesis approach using memristor crossbar,'' \emph{IEEE Transactions on Very Large Scale Integration (VLSI) Systems}, vol.~26, no.~2, pp. 355--366, 2018.

\bibitem{r17}
G.~Papandroulidakis, I.~Vourkas, A.~Abusleme, G.~C. Sirakoulis, and A.~Rubio, ``Crossbar-based memristive logic-in-memory architecture,'' \emph{IEEE Transactions on Nanotechnology}, vol.~16, no.~3, pp. 491--501, 2017.

\bibitem{r18}
Y.~Xi, B.~Gao, J.~Tang, A.~Chen, M.-F. Chang, X.~S. Hu, J.~V.~D. Spiegel, H.~Qian, and H.~Wu, ``In-memory learning with analog resistive switching memory: A review and perspective,'' \emph{Proceedings of the IEEE}, vol. 109, no.~1, pp. 14--42, 2021.

\bibitem{r2}
M.~Davies, N.~Srinivasa, T.-H. Lin, G.~Chinya, Y.~Cao, S.~H. Choday, G.~Dimou, P.~Joshi, N.~Imam, S.~Jain, Y.~Liao, C.-K. Lin, A.~Lines, R.~Liu, D.~Mathaikutty, S.~McCoy, A.~Paul, J.~Tse, G.~Venkataramanan, Y.-H. Weng, A.~Wild, Y.~Yang, and H.~Wang, ``Loihi: A neuromorphic manycore processor with on-chip learning,'' \emph{IEEE Micro}, vol.~38, no.~1, pp. 82--99, 2018.

\bibitem{r3}
F.~Akopyan, J.~Sawada, A.~Cassidy, R.~Alvarez-Icaza, J.~Arthur, P.~Merolla, N.~Imam, Y.~Nakamura, P.~Datta, G.-J. Nam, B.~Taba, M.~Beakes, B.~Brezzo, J.~B. Kuang, R.~Manohar, W.~P. Risk, B.~Jackson, and D.~S. Modha, ``Truenorth: Design and tool flow of a 65 mw 1 million neuron programmable neurosynaptic chip,'' \emph{IEEE Transactions on Computer-Aided Design of Integrated Circuits and Systems}, vol.~34, no.~10, pp. 1537--1557, 2015.

\bibitem{r4}
S.~Furber and A.~Brown, ``Biologically-inspired massively-parallel architectures - computing beyond a million processors,'' in \emph{2009 Ninth International Conference on Application of Concurrency to System Design}, 2009, pp. 3--12.

\bibitem{r5}
S.~Furber, S.~Temple, and A.~Brown, ``High-performance computing for systems of spiking neurons,'' in \emph{AISB’06 workshop on GC5: Architecture of Brain and Mind}, vol.~2, 2006, pp. 29--36.

\bibitem{r7}
J.~Schemmel, D.~Br{\"u}derle, A.~Gr{\"u}bl, M.~Hock, K.~Meier, and S.~Millner, ``A wafer-scale neuromorphic hardware system for large-scale neural modeling,'' in \emph{2010 IEEE International Symposium on Circuits and Systems (ISCAS)}.\hskip 1em plus 0.5em minus 0.4em\relax IEEE, 2010, pp. 1947--1950.

\bibitem{r6}
A.~Z. Foshie, N.~N. Chakraborty, J.~J. Murray, T.~J. Fowler, M.~S.~A. Shawkat, and G.~S. Rose, ``A multi-context neural core design for reconfigurable neuromorphic arrays,'' in \emph{IEEE Computer Society Annual Symposium on VLSI (ISVLSI)}, July 2021, pp. 67--72.

\bibitem{beckmann2020towards}
K.~Beckmann, W.~Olin-Ammentorp, G.~Chakma, S.~Amer, G.~S. Rose, C.~Hobbs, J.~V. Nostrand, M.~Rodgers, and N.~C. Cady, ``Towards synaptic behavior of nanoscale reram devices for neuromorphic computing applications,'' \emph{ACM Journal on Emerging Technologies in Computing Systems (JETC)}, vol.~16, no.~2, pp. 1--18, 2020.

\bibitem{liehr2019fabrication}
M.~Liehr, J.~Hazra, K.~Beckmann, W.~Olin-Ammentorp, N.~Cady, R.~Weiss, S.~Sayyaparaju, G.~Rose, and J.~Van~Nostrand, ``Fabrication and performance of hybrid reram-cmos circuit elements for dynamic neural networks,'' in \emph{Proceedings of the International Conference on Neuromorphic Systems}, 2019, pp. 1--4.

\bibitem{ryan}
R.~J. Weiss, ``Hardware for memristive neuromorphic systems with reliable programming and online learning,,'' The University of Tennessee, Knoxville, December 2022.

\bibitem{Programming_Scheme}
J.~Woo, K.~Moon, J.~Song, M.~Kwak, J.~Park, and H.~Hwang, ``Optimized programming scheme enabling linear potentiation in filamentary hfo2 rram synapse for neuromorphic systems,'' \emph{IEEE Transactions on Electron Devices}, vol.~63, no.~12, pp. 5064--5067, 2016.

\bibitem{binaryMemristor}
\BIBentryALTinterwordspacing
J.~Bill and R.~Legenstein, ``A compound memristive synapse model for statistical learning through stdp in spiking neural networks,'' \emph{Frontiers in Neuroscience}, vol.~8, 2014. [Online]. Available: \url{https://www.frontiersin.org/article/10.3389/fnins.2014.00412}
\BIBentrySTDinterwordspacing

\bibitem{r19}
H.~Kim, M.~P. Sah, C.~Yang, T.~Roska, and L.~O. Chua, ``Neural synaptic weighting with a pulse-based memristor circuit,'' \emph{IEEE Transactions on Circuits and Systems I: Regular Papers}, vol.~59, no.~1, pp. 148--158, 2011.

\bibitem{r20}
S.~Mukhopadhyay, H.~Mahmoodi, and K.~Roy, ``Modeling of failure probability and statistical design of sram array for yield enhancement in nanoscaled cmos,'' \emph{IEEE Transactions on Computer-Aided Design of Integrated Circuits and Systems}, vol.~24, no.~12, pp. 1859--1880, 2005.

\bibitem{model_glsvlsi}
H.~Das, M.~Rathore, R.~Febbo, M.~Liehr, N.~C. Cady, and G.~S. Rose, ``Rfam: Reset-failure-aware-model for hfo2-based memristor to enhance the reliability of neuromorphic design,'' in \emph{Proceedings of the Great Lakes Symposium on VLSI 2023}, 2023, pp. 281--286.

\bibitem{65}
K.~Beckmann, J.~Holt, J.~Capulong, S.~Lombardo, N.~C. Cady, and J.~Van~Nostrand, ``Reliability of fully-integrated nanoscale reram/cmos combinations as a function of on-wafer current control,'' in \emph{2014 IEEE International Integrated Reliability Workshop Final Report (IIRW)}, 2014, pp. 159--162.

\bibitem{r10}
Y.~M. Ding, D.~D. Misra, and P.~Srinivasan, ``Flicker noise performance on thick and thin oxide finfets,'' \emph{IEEE Transactions on Electron Devices}, vol.~64, no.~5, pp. 2321--2325, 2017.

\bibitem{psb:18:ten}
\BIBentryALTinterwordspacing
J.~S. Plank, C.~D. Schuman, G.~Bruer, M.~E. Dean, and G.~S. Rose, ``The {TENNLab} exploratory neuromorphic computing framework,'' \emph{IEEE Letters of the Computer Society}, vol.~1, no.~2, pp. 17--20, July-Dec 2018. [Online]. Available: \url{https://doi.ieeecomputersociety.org/10.1109/LOCS.2018.2885976}
\BIBentrySTDinterwordspacing

\bibitem{smp:20:eons}
C.~D. Schuman, J.~P. Mitchell, R.~M. Patton, T.~E. Potok, and J.~S. Plank, ``Evolutionary optimization for neuromorphic systems,'' in \emph{NICE: Neuro-Inspired Computational Elements Workshop}, 2020.

\bibitem{prs:19:tsl}
\BIBentryALTinterwordspacing
J.~S. Plank, C.~Rizzo, K.~Shahat, G.~Bruer, T.~Dixon, M.~Goin, G.~Zhao, J.~Anantharaj, C.~D. Schuman, M.~E. Dean, G.~S. Rose, N.~C. Cady, and J.~{Van Nostrand}, ``The {TENNLab} suite of {LIDAR}-based control applications for recurrent, spiking, neuromorphic systems,'' in \emph{44th Annual GOMACTech Conference}, Albuquerque, March 2019. [Online]. Available: \url{http://neuromorphic.eecs.utk.edu/raw/files/publications/2019-Plank-Gomac.pdf}
\BIBentrySTDinterwordspacing

\bibitem{r23}
\BIBentryALTinterwordspacing
A.~Z. Foshie, C.~Rizzo, H.~Das, C.~Zheng, J.~S. Plank, and G.~S. Rose, ``Benchmark comparisons of spike-based reconfigurable neuroprocessor architectures for control applications,'' in \emph{Proceedings of the Great Lakes Symposium on VLSI 2022}, ser. GLSVLSI '22.\hskip 1em plus 0.5em minus 0.4em\relax New York, NY, USA: Association for Computing Machinery, 2022, p. 383–386. [Online]. Available: \url{https://doi.org/10.1145/3526241.3530381}
\BIBentrySTDinterwordspacing

\bibitem{asuncion2007uci}
A.~Asuncion and D.~Newman, ``Uci machine learning repository,'' 2007.

\bibitem{ref1}
Y.~Ge, L.~Xu, M.~Chen, Y.~Yang, and F.~Wu, ``A mixed-signal charge-mode synapse circuit with good linearity and low power consumption,'' in \emph{2018 2nd IEEE Advanced Information Management,Communicates,Electronic and Automation Control Conference (IMCEC)}, 2018, pp. 1077--1080.

\bibitem{ref2}
\BIBentryALTinterwordspacing
L.~Chen, Z.-Y. He, T.-Y. Wang, Y.-W. Dai, H.~Zhu, Q.-Q. Sun, and D.~W. Zhang, ``Cmos compatible bio-realistic implementation with ag/hfo2-based synaptic nanoelectronics for artificial neuromorphic system,'' \emph{Electronics}, vol.~7, no.~6, 2018. [Online]. Available: \url{https://www.mdpi.com/2079-9292/7/6/80}
\BIBentrySTDinterwordspacing

\bibitem{ref3}
\BIBentryALTinterwordspacing
S.~Rehman, M.~F. Khan, S.~Aftab, H.~Kim, J.~Eom, and D.-k. Kim, ``Thickness-dependent resistive switching in black phosphorus cbram,'' \emph{J. Mater. Chem. C}, vol.~7, pp. 725--732, 2019. [Online]. Available: \url{http://dx.doi.org/10.1039/C8TC04538K}
\BIBentrySTDinterwordspacing

\bibitem{ref4}
\BIBentryALTinterwordspacing
S.~Sonde, B.~Chakrabarti, Y.~Liu, K.~Sasikumar, J.~Lin, L.~Stan, R.~Divan, L.~E. Ocola, D.~Rosenmann, P.~Choudhury, K.~Ni, S.~K. R.~S. Sankaranarayanan, S.~Datta, and S.~Guha, ``Silicon compatible sn-based resistive switching memory,'' \emph{Nanoscale}, vol.~10, pp. 9441--9449, 2018. [Online]. Available: \url{http://dx.doi.org/10.1039/C8NR01540F}
\BIBentrySTDinterwordspacing

\bibitem{ref5}
\BIBentryALTinterwordspacing
Y.~Ahn, H.~W. Shin, T.~H. Lee, W.-H. Kim, and J.~Y. Son, ``Effects of a nb nanopin electrode on the resistive random-access memory switching characteristics of nio thin films,'' \emph{Nanoscale}, vol.~10, pp. 13\,443--13\,448, 2018. [Online]. Available: \url{http://dx.doi.org/10.1039/C8NR02986E}
\BIBentrySTDinterwordspacing

\bibitem{ref6}
S.~De, F.~Müller, H.-H. Le, M.~Lederer, Y.~Raffel, T.~Ali, D.~Lu, and T.~Kämpfe, ``Read-optimized 28nm hkmg multibit fefet synapses for inference-engine applications,'' \emph{IEEE Journal of the Electron Devices Society}, vol.~10, pp. 637--641, 2022.

\bibitem{r21}
M.~Hu, C.~E. Graves, C.~Li, Y.~Li, N.~Ge, E.~Montgomery, N.~Davila, H.~Jiang, R.~S. Williams, J.~J. Yang \emph{et~al.}, ``Memristor-based analog computation and neural network classification with a dot product engine,'' \emph{Advanced Materials}, vol.~30, no.~9, p. 1705914, 2018.

\bibitem{r22}
R.~Weiss, H.~Das, N.~N. Chakraborty, and G.~S. Rose, ``Stdp based online learning for a current-controlled memristive synapse,'' in \emph{2022 IEEE 65th International Midwest Symposium on Circuits and Systems (MWSCAS)}, 2022, pp. 1--4.

\bibitem{Gangotree}
G.~Chakma, M.~M. Adnan, A.~R. Wyer, R.~Weiss, C.~D. Schuman, and G.~S. Rose, ``Memristive mixed-signal neuromorphic systems: Energy-efficient learning at the circuit-level,'' \emph{IEEE Journal on Emerging and Selected Topics in Circuits and Systems}, vol.~8, no.~1, pp. 125--136, 2018.

\bibitem{TCAS}
H.~Das, R.~D. Febbo, S.~N.~B. Tushar, N.~N. Chakraborty, M.~Liehr, N.~C. Cady, and G.~S. Rose, ``An efficient and accurate memristive memory for array-based spiking neural networks,'' \emph{IEEE Transactions on Circuits and Systems I: Regular Papers}, pp. 1--12, 2023.

\end{thebibliography}

\vspace{11pt}
\bf{}\vspace{-33pt}
\begin{IEEEbiography}[{\includegraphics[width=1in,height=1.25in,clip,keepaspectratio]{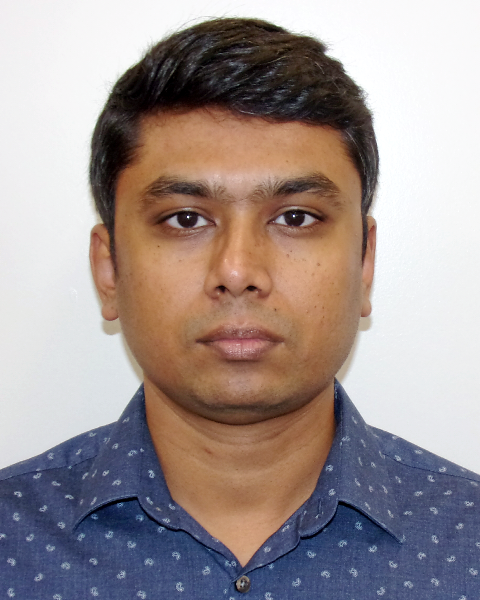}}]{Hritom Das} (Member, IEEE) received the B.Sc. degree in electrical and electronic engineering from American International University-Bangladesh, Dhaka, Bangladesh, the M.Sc. degree in electronic engineering from Kyungpook National University, Daegu, South Korea, and the Ph.D. degree in electrical and computer engineering from North Dakota State University, Fargo, North Dakota, in 2012, 2015, and 2020 respectively. He was a visiting Assistant Professor with the department of electrical and computer engineering at the University of South Alabama, Mobile, AL, USA. Currently, he is a Post-Doctoral Research Associate with the department of electrical engineering and computer science at The University of Tennessee, Knoxville, TN, USA. His research interest includes low-power VLSI circuit design, optimization, and testing. He is also exploring machine learning implementation on traditional electronics.  In addition, he is working on neuromorphic system design and optimization.
\end{IEEEbiography}

 
\begin{IEEEbiography}[{\includegraphics[width=1in,height=1.25in,clip,keepaspectratio]{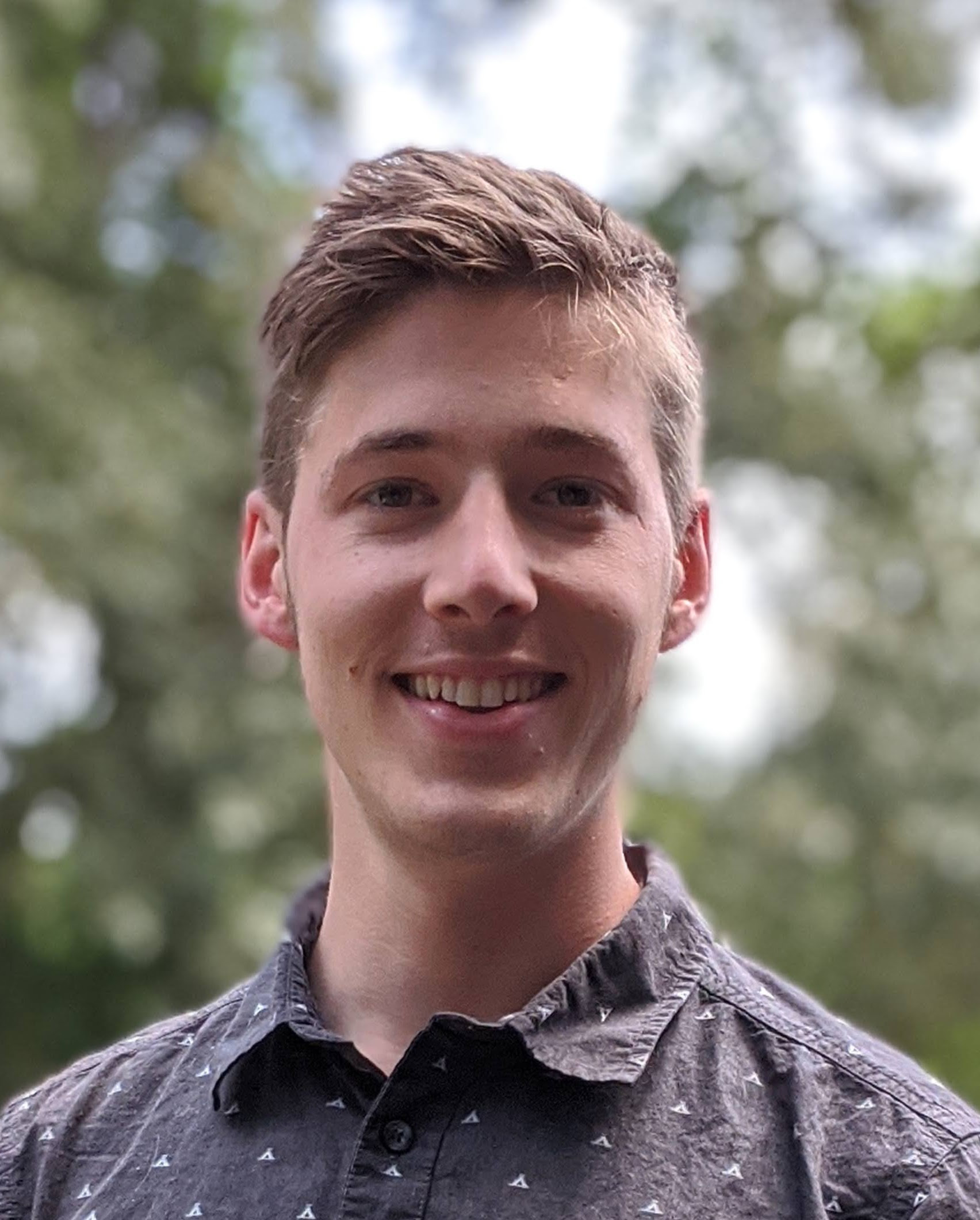}}] {Rocco Febbo} is currently enrolled at University of Tennessee, Knoxville (UTK) as a Ph.D student. He received his B.S in electrical engineering from UTK in 2020. His research interests include working with memristive devices to design and test low power neuromorphic circuits as well as designing digital systems for neuromorphic computing.
\end{IEEEbiography}

 \vspace{-1.1cm}
\begin{IEEEbiography}[{\includegraphics[width=1in,height=1.25in,clip,keepaspectratio]{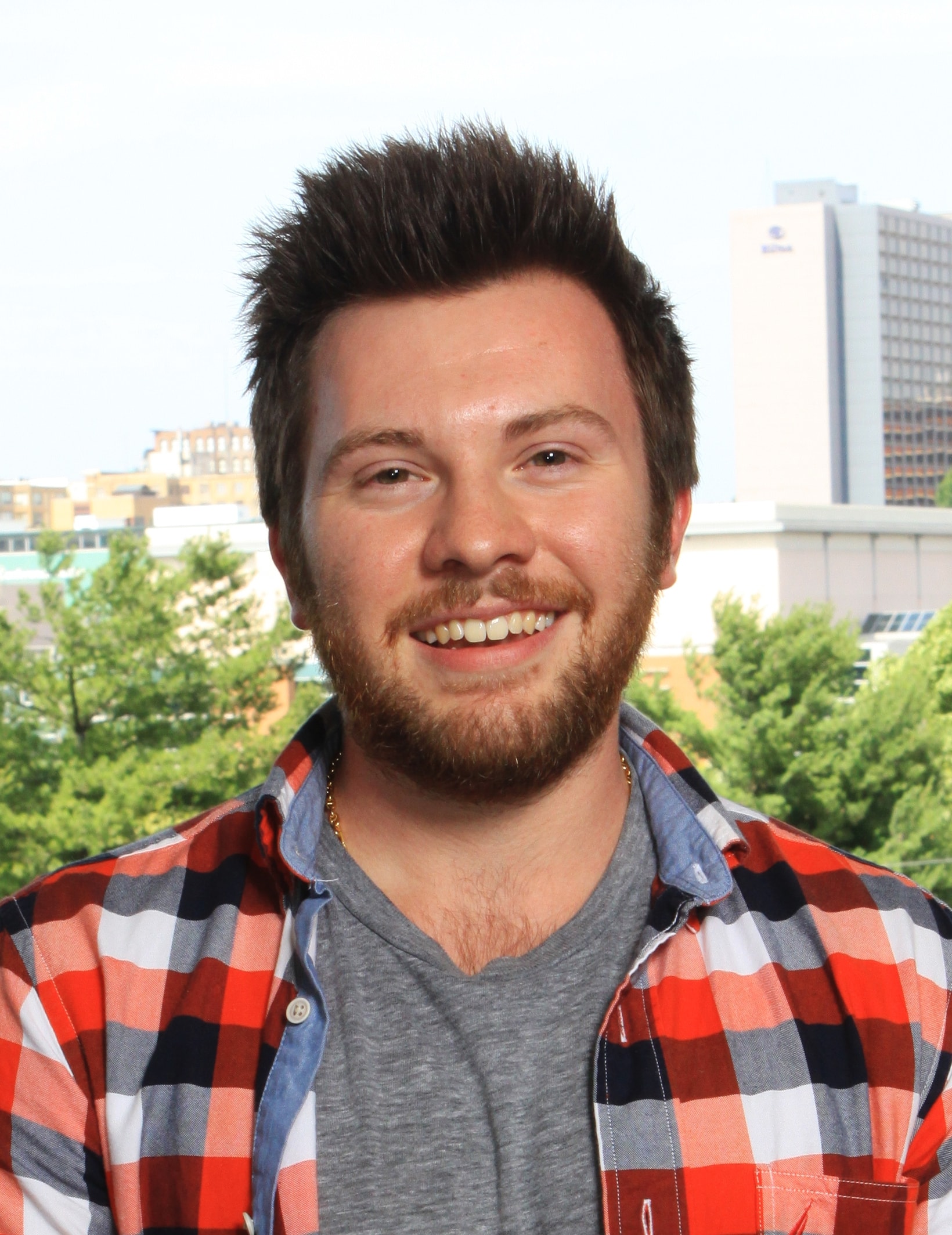}}] {Charles P. Rizzo} is currently enrolled at University of Tennessee, Knoxville (UTK) as a Ph.D student. He received his B.S in computer science from UTK in 2019 and his M.S in computer science from UTK in 2021. His research interests include working with event-based vision sensors and applying neuromorphic solutions to control and classification tasks using event-based vision.
\end{IEEEbiography}
\vspace{-1.1cm}
\begin{IEEEbiography}[{\includegraphics[width=1in,height=1.25in,clip,keepaspectratio]{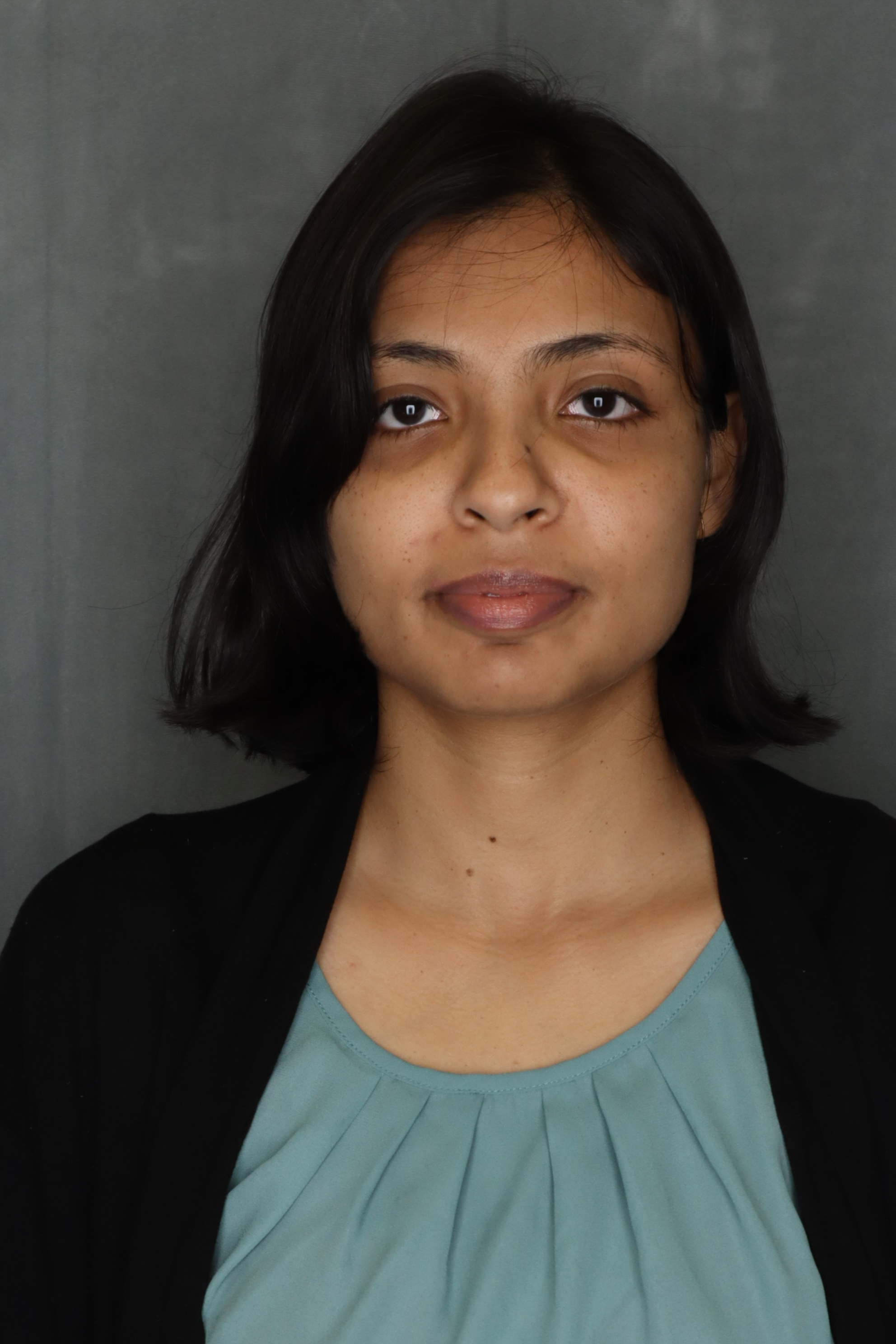}}] {Nishith N. Chakraborty} (Graduate Student Member, IEEE) is currently enrolled at the University of Tennessee, Knoxville (UTK) as a Ph.D student. She received her B.Sc. degree in electrical and electronic engineering from Bangladesh University of Engineering and Technology, and MS in electrical engineering from University of California, Riverside. Her research interest includes analog mixed-signal neuromorphic circuits design and optimization, low-power VLSI circuits and memristor-based circuit design.
\end{IEEEbiography}

\vspace{-10cm}
\begin{IEEEbiography}[{\includegraphics[width=1in,height=1.25in,clip,keepaspectratio]{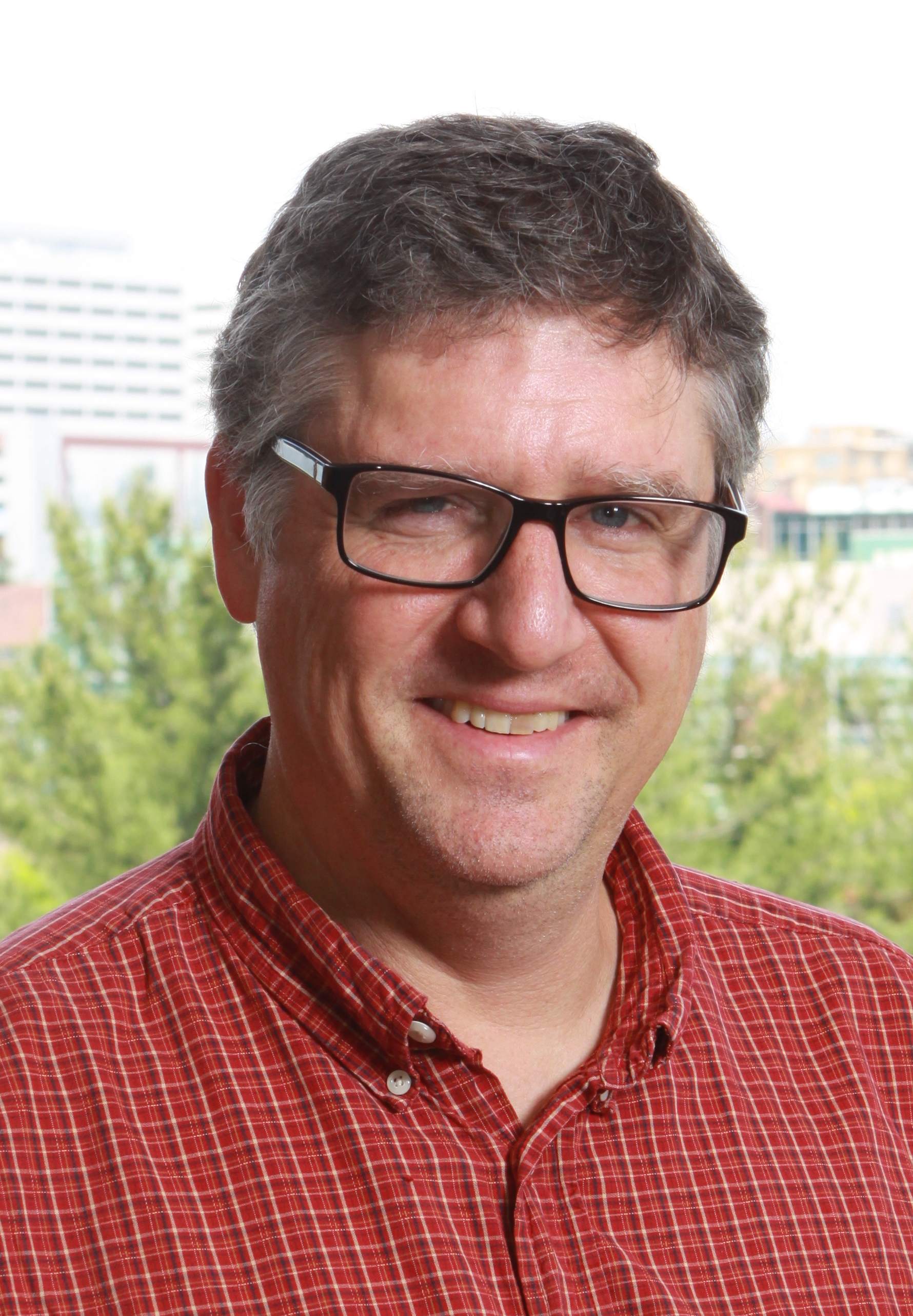}}] {James S. Plank} (Senior Member, IEEE) James S. Plank received his B.S. from Yale University in 1988 and his
Ph.D. from Princeton University in 1993. He is a Professor in the
Electical Engineering and Computer Science department at the
University of Tennessee, where he has been since 1993. His research interests are in neuromorphic computing, specifically systems, simulators, applications and algorithms.
He is a past Associate Editor of IEEE
Transactions on Parallel and Distributed Computing, and a Senior Member of
IEEE Computer Society.
\end{IEEEbiography}

\vspace{-10cm}
\begin{IEEEbiography}[{\includegraphics[width=1in,height=1.25in,clip,keepaspectratio]{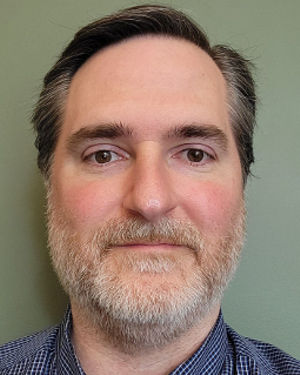}}] {Garrett S. Rose} (Senior Member, IEEE) received
the B.S. degree in computer engineering from Virginia Polytechnic Institute and State University, Blacksburg, VA, USA, in 2001, and the M.S. and Ph.D. degrees in electrical engineering from the University of Virginia, Charlottesville, VA, USA,in 2003 and 2006, respectively. His Ph.D. dissertation was on the topic of circuit design methodologies for molecular electronic circuits and computingv architectures.He is currently a Professor with the Department of Electrical Engineering and Computer Science, University of
Tennessee, Knoxville, TN, USA, where his work is focused on research in the areas of nanoelectronic circuit design, neuromorphic computing, and hardware security. From June 2011 to July 2014, he was with the Air Force Research Laboratory, Information Directorate, Rome, NY, USA. From August 2006 to May 2011, he was an Assistant Professor with the Department of Electrical and Computer Engineering, Polytechnic Institute of New York University, Brooklyn, NY, USA. From May 2004 to August 2005, he was with MITRE Corporation, McLean, VA, USA, involved in the design and simulation of
nanoscale circuits and systems. His research interests include low-power circuits, system-on-chip design, trusted hardware, and developing VLSI design methodologies for novel nanoelectronic technologies 
\end{IEEEbiography}

\end{document}